\documentclass[journal,english]{IEEEtran}

\usepackage{comment}

\usepackage[T1]{fontenc}
\usepackage{babel}
\usepackage{calc}
\usepackage{amsthm}
\usepackage{epsfig}
\usepackage{times}
\usepackage{float}
\usepackage{afterpage}
\usepackage{amsmath}
\usepackage{amstext,cite}
\usepackage{amssymb,bm}
\usepackage{latexsym}
\usepackage{color}
\usepackage{graphicx}
\usepackage{amsmath}
\usepackage{multirow}
\usepackage{graphicx}
\usepackage{pstricks}
\usepackage{booktabs}
\usepackage{multicol}
\usepackage{lipsum}
\usepackage{epstopdf}
\usepackage{mathrsfs,mathtools}
\usepackage[unicode=true,
 bookmarks=true,
 bookmarksnumbered=true,
 bookmarksopen=true,
 bookmarksopenlevel=1,
 breaklinks=false,
 pdfborder={0 0 0},
 backref=false,
 colorlinks=false]
{hyperref}
\hypersetup{pdftitle={Your Title},
 pdfauthor={Your Name},
 pdfpagelayout=OneColumn,
 pdfnewwindow=true,
 pdfstartview=XYZ,
 plainpages=false}
\usepackage{soul}
\usepackage{cancel}

\makeatletter

\newcommand{\Ac}{\mathcal{A}}
\newcommand{\Bc}{\mathcal{B}}
\newcommand{\Dc}{\mathcal{D}}
\newcommand{\Jc}{\mathcal{J}}
\newcommand{\Kc}{\mathcal{K}}
\newcommand{\Wc}{\mathcal{W}}
\newcommand{\Sc}{\mathcal{S}}
\newcommand{\Pc}{\mathcal{P}}
\newcommand{\Xc}{\mathcal{X}}
\newcommand{\Yq}{s_{q}} 

\newtheorem{thm}{Theorem} 
\newtheorem{cor}{Corollary}
\newtheorem{lem}{Lemma}

\newtheorem{rem}{Remark}
\newtheorem{example}{Example}
\newtheorem{defn}{Definition}

\begin{document}

\title{An Index Coding Approach to Caching \\ with Uncoded Cache Placement}

\author{
Kai~Wan,~\IEEEmembership{Member,~IEEE,} 
Daniela~Tuninetti,~\IEEEmembership{Senior~Member,~IEEE,}
and~Pablo~Piantanida,~\IEEEmembership{Senior~Member,~IEEE}
\thanks{The results of this paper were presented in parts at the following conferences:
   the IEEE 2016 Information Theory Workshop, Cambridge, UK,
   the IEEE 2016 International Symposium on Information Theory, Barcelona, Spain, and
   the IEEE 2017 Information Theory and Applications Workshop, San Diego, CA USA. 
   The results were also made available  online as of Nov. 2015 as follows:  
   arXiv:1511.02256,
   arXiv:1601.06383, and
   arXiv:1702.07265.}
\thanks{
K.~Wan was with Laboratoire de Signaux et Syst\`emes (L2S, UMR8506), CentraleSup{\'e}lec-CNRS-Universit{\'e} Paris-Sud, 91192 Gif-sur-Yvette, France.
He is now with  Technische Universit\"at Berlin, Berlin, Germany (email: kai.wan@tu-berlin.de).}
\thanks{
D.~Tuninetti is with the Electrical and Computer Engineering Department, University of Illinois at Chicago, Chicago, IL 60607 USA (e-mail: danielat@uic.edu).} 
\thanks{
P.~Piantanida is with CentraleSup{\'e}lec--French National Center for Scientific Research (CNRS)--Universit{\'e} Paris-Sud, 3 Rue Joliot-Curie, F-91192 Gif-sur-Yvette, France, and with Montreal Institute for Learning Algorithms (MILA) at Universit{\'e} de Montr{\'e}al, 2920 Chemin de la Tour, Montr{\'e}al, QC H3T 1N8, Canada (e-mail: pablo.piantanida@centralesupelec.fr).}
\thanks{
The work of K.~Wan was  supported by Labex DigiCosme (project ANR11LABEX0045DIGICOSME) operated by ANR as part of the program ``Investissement d'Avenir'' Idex ParisSaclay (ANR11IDEX000302).
The work of D. Tuninetti was supported in part by the National Science Foundation under award number 1527059.
The work of P.~Piantanida was supported by the European Commission's Marie Sklodowska-Curie Actions (MSCA), through the Marie Sklodowska-Curie IF (H2020-MSCAIF-2017-EF-797805-STRUDEL).}
}

\maketitle

\begin{abstract}
Caching is an efficient way to reduce network traffic congestion during peak hours, by storing some content at the user's local cache memory, even without knowledge of user's later demands. Maddah-Ali and Niesen proposed a two-phase (placement phase and delivery phase) coded caching strategy for broadcast channels with cache-aided users.
This paper investigates the same model under the constraint that content is placed uncoded within the caches, that is, when bits of the files are simply copied within the caches.
When the cache contents are uncoded and the users' demands are revealed, the caching problem can be connected to an index coding problem. This paper focuses on deriving fundamental performance limits for the caching problem by using tools for the index coding problem that were either known or are newly developed in this work. 

First, a converse bound for the caching problem under the constraint of uncoded cache placement is proposed based on the ``acyclic index coding converse bound.''  This converse bound is proved to be achievable by the Maddah-Ali and Niesen's scheme when the number of files is not less than the number of users, and by a newly derived index coding achievable scheme otherwise.  
The proposed index coding achievable scheme is based on distributed source coding and strictly improves on the widely used  ``composite (index) coding'' achievable bound and its improvements, and is of independent interest.
%

An important consequence of the findings of this paper is that advancements on the coded caching problem posed by Maddah-Ali and Niesen are thus only possible by considering strategies with coded placement phase. A recent work  by Yu \emph{et al} has however shown that coded cache placement can at most half the network load compared to the results presented in this paper.
\end{abstract}



\section{Introduction}
\label{sec:intro}


\IEEEPARstart{N}{etworks} have ``rush hours,'' with peak traffic hours where traffic is high and the network performance suffers, and off-peak times, where traffic is low. Caching is an effective method to smooth out network traffic during peak times. 
In cache-aided networks, some content is locally stored into the users' local cache memory during off-peak hours in the hope that the pre-stored content will be required during peak hours. When this happens, content is retrieved locally thereby reducing the communication load, or number of transmissions, from the server to the users.

In this paper, we study the fundamental performance limits of cache-aided broadcast systems (also known as single bottleneck shared-link model) by following the model originally proposed by Maddah-Ali and Niesen (MAN) in their seminar works~\cite{dvbt2fundamental,decentralizedcoded}.  We focus on the practically relevant case when content is stored uncoded in the local caches, in which case the caching problem can be related to the Index Coding (IC) problem. 
Although the connection between caching and IC is well known~\cite{dvbt2fundamental,orderrandomJi2017}, to the best of our knowledge, IC results have not been used to characterize the performance of the caching problem in the literature prior to our first work~\cite{ontheoptimality}, which was publicly available online since November 2015.
\emph{This paper's main contribution is to leverage both known and hereby newly derived results for the IC problem to determine the fundamental limits of cache-aided systems with uncoded cache placement.} 
Since the publication of our work, a significant body of work on cache-aided systems has focused on characterizing the ultimate performance limits of caching schemes under the constraint of uncoded cache placement as we did  in~\cite{ontheoptimality, ourisitinnerbound}.
A non-exhaustive list of such works includes Device-to-Device systems~\cite{ourd2dpaper}, coded caching systems with heterogeneous cache sizes~\cite{yener2018D2Dhetero}, topological coded caching systems~\cite{wan2017combinationouter}, coded caching systems with shared caches~\cite{parrinello2018sharedcache}, coded data shuffling~\cite{approximatelyDS2018Attia}, etc.

\subsection{Past Work}
\label{sec:intro:pastall}

\paragraph*{MAN's work}
In~\cite{dvbt2fundamental}, Maddah-Ali and Niesen proposed a coded caching scheme that utilizes an uncoded combinatorial cache construction in the placement phase and a binary linear network code in the delivery phase, where content in the caches is stored in a coordinated manner.
The key observation is that well designed packets in the delivery phase are able to simultaneously satisfy many users at once, thus providing a ``global caching gain'' that scales with the total cache size in the network, in addition to the well known ``local caching gain'' that only depends on the amount of local cache at each user. In~\cite{dvbt2fundamental}, Maddah-Ali and Niesen analytically showed that the load of their proposed scheme is to within a factor of $12$ of a cut-set-type converse bound, but it was noted in~\cite{wang2016anewconverse} that it numerically appeared to be optimal to within a factor $4.7$.
The scheme in~\cite{dvbt2fundamental} has been improved in many ways; examples of schemes with uncoded cache placement are~\cite{cachingwithlargeumberusers, ontheoptimality, ourisitinnerbound, exactrateuncoded}, while with coded cache placement are~\cite{codedcachingviaintef, smallbufferusers, kuserstwofiles, improveddelivery, jesus2017fundamental}.

\paragraph*{Converse bounds for any cache placement}
In~\cite{improvedlower, criticaldatabase, ISIT2015outerbound, demandtype,improvedconverse2017Wang
}, converse bounds tighter than cut-set bound provided in~\cite{dvbt2fundamental} and valid for any type of cache placement were proposed.
%
An improved converse bound compared to the cut-set bound was given in~\cite[Theorem 1]{criticaldatabase}, which was used to show that the effectiveness of caching becomes small when the number of files becomes comparable to the square of the number of users.
%
An algorithm 
that generalizes~\cite{criticaldatabase} was proposed in~\cite{improvedlower} to generate  lower  bounds on $\alpha R + \beta M$, for positive integers $(\alpha,\beta)$ and where $R$ is the load and $M$ the cache size, and used to show that the achievable load in~\cite{dvbt2fundamental} is optimal to within a factor $4$. 
%
Another converse bound was obtained in~\cite{ISIT2015outerbound} by leveraging~\cite[Theorem 17.6.1]{CoverThomas2ndEdition} as a `symmetrization argument' over demands and used to show that the achievable load in~\cite{dvbt2fundamental} is optimal to within a factor $8$; the converse bound applies to the case where users can request multiple files from the server as well. 
Inspired by converse results for caching systems over general degraded broadcast channels~\cite{benefitofcache2017Shirin}, the authors in~\cite{improvedconverse2017Wang} 
proposed achievable and converse bounds for the worst-case and the average loads that are to within a multiplicative gap of $2.315$ and $2.507$, respectively. 
%
An approach based on solving a linear program derived from the sub-modularity of entropy and simplified by leveraging certain inherent symmetries in the caching problem was put forth in~\cite{demandtype} as a means to computationally generate converse bounds; the approach allowed to solve the case of $K=2$ users and any number of files $N$, and gives at present the tightest bounds for problems with small $K$ and $N$ (beyond which the computational approach becomes practically unfeasible).

\paragraph*{Converse bounds for uncoded cache placement}
 A different line of work that was initiated with our work in~\cite{ontheoptimality}, where we asked the question of what would be the ultimate  performance limit of cache-aided systems if one restricts the placement phase to be uncoded. In~\cite{ontheoptimality}, we studied such a setting from the lens of IC. 
The IC had been connected to coded caching earlier in~\cite[p.~2865, Section~VIII.A]{dvbt2fundamental}: ``
[...] Now, for fixed \emph{uncoded} content placement and for fixed demands, the delivery phase of the caching problem induces a so-called IC problem. However, it is important to realize that the caching problem actually consists of exponentially many parallel such IC problems, one for each of the $N^K$ possible user demands. Furthermore, the IC problem itself is computationally hard to solve even only approximately. The main contribution of this paper is to design the content placement such that each of the exponentially many parallel IC problems has simultaneously an efficient and analytical solution.''  In~\cite{ontheoptimality}, we analyzed the performance of the caching problem for fixed uncoded content placement and for fixed demands as an IC; by leveraging a known IC converse bound, and by carefully picking certain user demands, we explicitly characterized the worst case load as a function of certain parameters of the placement phase, which we (Fourier Motzkin) eliminated to find a closed form expression for the optimal load. This paper is the long, journal version, of our series of conference works that stared with~\cite{ontheoptimality}. 

The exact memory-load tradeoff for cache-aided systems under the constraint of uncoded cache placement was characterized in~\cite{exactrateuncoded}. 
The converse bound in~\cite{exactrateuncoded} is derived by a genie-based idea (instead of directly leveraging the acyclic IC converse bound as we did in~\cite{ontheoptimality,ourisitinnerbound}), which is equivalent to our approach here -- see the first remark Section~\ref{sub:converse remark}. 
The genie-based idea was also extended to the case of average load (as opposed to worst case load) with uniform independent and identically distributed demands. Our IC-based approach also extends to the same average lead setting -- see the second remark in Section~\ref{sub:converse remark}.

\paragraph*{Optimality for uncoded cache placement}
By enhancing the cut-set-type converse bound by an additional non-negative term, the achievable load in~\cite{dvbt2fundamental} and its enhanced version in~\cite{exactrateuncoded} were proved to be optimal to within a factor $2$~\cite{yu2017characterizing}.  This is, to the best of our knowledge, the sharpest known multiplicative gap, and implies that coded cache placement can at most half the network load compared to the results presented in~\cite{ontheoptimality} (and in this paper) and in~\cite{exactrateuncoded}.

\paragraph*{Achievability and extensions}
Much work has gone into improving the MAN caching scheme, especially in the small cache size regime where coded cache placement can outperform uncoded placement (as already noted in~\cite{dvbt2fundamental
}). 
We shall not deal further into this line of work as it is not relevant to our work here. 

The MAN caching scheme has been generalized to account for
caching with nonuniform demands,
multi-demands,
shared caches,
distinct file sizes or distinct cache sizes,
online caching placement, 
hierarchical coded caching or other network topologies, 
device-to-device applications,
secrete caching schemes, 
cache-aided systems with finite file size, 
etc. 
The results for these important practical scenarios are not discuss here as they are not directly relevant for our work. We note that the connection between caching and IC can be also leveraged in these systems,  as we did for combination networks~\cite{novelouterwan2017}.

\subsection{Main Contributions} 
\label{sec:intro:ourresults}

\begin{table*}
\centering
\protect\caption{Summary of the paper contributions.}
\label{tab:contribution}
\begin{tabular}{|c|c|c|}
\hline 
Problem  
 & Proposed results 
 & Compared to the literature
\tabularnewline
\hline 
\hline 
Index coding 
  & Novel achievable scheme
  & Strictly outperform the state-of-the-art in~\cite{onthecapacityindex}
\tabularnewline
\hline 
Caching 
 & Novel converse bound for uncoded placement 
 & Originally proposed in~\cite{ontheoptimality,ourisitinnerbound},
   and later in~\cite{exactrateuncoded}
\tabularnewline
\hline 
Caching 
 & Our novel IC bound achieves our novel converse 
 & A generalized version of the scheme  in~\cite{exactrateuncoded}
from the viewpoint of index coding
\tabularnewline
\hline 
\end{tabular}
\end{table*}

Our main contributions, and how they compare to existing works, are summarized in Table~\ref{tab:contribution}. 
In a nut shell, we focus on cache-aided systems with uncoded placement 
and study their performance by drawing connections to the IC problem.  More precisely,
\begin{enumerate}

\item
\emph{Converse for cache-aided systems with uncoded cache placement.}   
In Section~\ref{sec:centralized converse}, by exploiting~\cite[Corollary 1]{onthecapacityindex} for the IC problem, we derive a converse bound for the load in centralized cache-aided systems with uncoded cache placement. 
We show that it matches the load in~\cite{dvbt2fundamental} when there are more files than users. 

\item
\emph{Novel IC achievable bound and its application to cache-aided systems.}
In Section~\ref{sec:centralized inner bound}, we propose an IC achievable bound based on Han's coding scheme~\cite{hanspaper}, Slepian-Wolf coding~\cite{slepianwolf} and non-unique decoding~\cite{elgamalkim}. This achievable scheme is shown to strictly outperform the composite (index) coding scheme, and is, to the best our knowledge, the best random coding achievable bound for the general IC problem to date. 
We then show that our novel IC scheme matches our converse bound for centralized cache-aided systems with uncoded cache placement when there are less files than users, and attains the same load as the scheme in~\cite{yu2017characterizing}.

\end{enumerate}

\subsection{Paper Outline}
\label{sec:intro:outline}
The rest of the report is organized as follows. 
The system models for centralized cache-aided systems and for IC, and their relationship, are introduced in Section~\ref{sec:model}, as well as the results on those two problems needed in later sections. 
In Section~\ref{sec:centralized converse}, we derive a converse bound under the constraint of uncoded cache placement. In Section~\ref{sec:centralized inner bound}, we introduce a novel IC achievable bound and use it to design a caching scheme that achieves the proposed converse bound for the caching problem. 
Section~\ref{sec:conclusions} concludes the paper. 
Some proofs may be found in Appendix.

Results for decentralized cache-aided systems\footnote{%
Cache-aided systems are divided into two classes, \emph{centralized}~\cite{dvbt2fundamental} and \emph{decentralized}~\cite{decentralizedcoded}, depending on whether users can coordinate during the placement phase. 
In centralized cache-aided systems, the users in the two phases of the caching scheme are assumed to be the same;  therefore, coordination among users is possible in the placement phase. 
In practice, 
for example due to users' mobility, a user may be connected to a server during its placement phase but to a different one during its delivery phase;
in such scenarios, coordination among users during the placement phase is thus not possible.
}
under the constraint of uncoded cache placement are not reported here because they follow from the same line of reasoning as those for centralized systems as detailed in~\cite{kai-defense} -- see also the third remark in Section~\ref{sub:achiev remark}. 

\section{System Models and Some Known Results}
\label{sec:model}

\subsection{Notation}
\label{sec:model:notation}
Calligraphic symbols denote sets; 
symbols in bold font denote vectors;
$|\cdot|$ is used to represent the cardinality of a set or the length of a file in bits;
we let 
$\mathcal{A\setminus B}\vcentcolon=\left\{ x\in\Ac|x\notin\mathcal{B}\right\}$,
$[a:b:c]\vcentcolon=\{a,a+b,a+2b,...,c\}$,
$[a:c] = [a:1:c]$ and $[n]=[1:n]$;
the bit-wise XOR operation between binary vectors is indicated by $\oplus$;
for two integers $x$ and $y$, we let $\binom{x}{y}=0$ if $x<y$ or $x\leq0$.

\subsection{The Centralized Caching Problem: Definition} 
\label{sec:model:caching}

\begin{figure}
\centering
\includegraphics[scale=0.4]{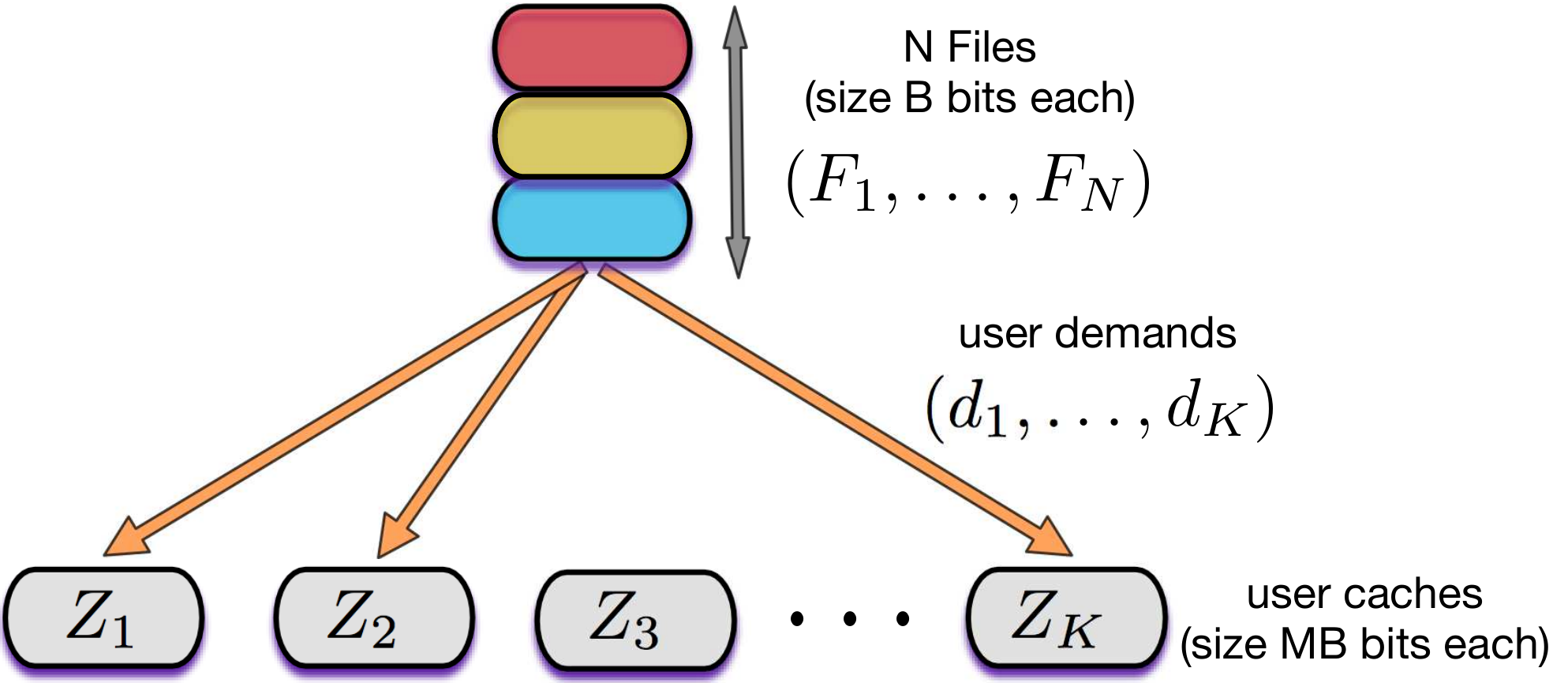}
\caption{\small 
A centralized cache-aided system where a server with $N$~files of size $B$~bits is connected to $K$~users equipped with a cache of size $MB$ bits. 
}
\label{fig:caching}
\end{figure}

The information-theoretic formulation of the centralized coded caching problem in Fig.~\ref{fig:caching}, as originally formulated by Maddah-Ali and Niesen in~\cite{dvbt2fundamental}, is as follows.
\begin{itemize}

\item
The system comprises a server with $N$ independent files, denoted by $(F_{1},F_{2},\dots,F_{N})$, and $K$ users connected to it through an error-free link. Each file has $B$ independent and equally likely bits. 

\item
In the placement phase, user $k\in[K]$ stores content from the $N$ files in its cache of size $MB$ bits without knowledge of later demands, where $M\in[0,N]$. We denote the content in the cache of user $k\in[K]$ by $Z_{k}=\phi_{k}(F_{1},\ldots,F_{N})$, where 
\begin{align}
\phi_{k} : 
[0:1]^{NB}\rightarrow[0:1]^{\left\lfloor MB \right\rfloor }, \ \forall k\in[K].
\label{eq:caching placement def}
\end{align}
We also denote by $\mathbf{Z}\vcentcolon=(Z_{1},\ldots,Z_{K})$ the content of all the caches.

\item
In the delivery phase, each user demands one file and the demand vector $\mathbf{d}\vcentcolon=(d_{1},d_{2},\dots,d_{K})$, where $d_{k}\in[N]$ corresponds to the file demanded by user $k\in[K]$, is revealed to the server and all users. 
Given $(\mathbf{Z},\mathbf{d})$, the server broadcasts the message $X_{\mathbf{d}} = \psi(F_{1},\ldots,F_{N},\mathbf{d})$, where 
\begin{align}
\psi : 
[0:1]^{NB} 
\times[N]^{K} \rightarrow[0:1]^{\left\lceil RB\right\rceil}.
\label{eq:caching enc def}
\end{align}

\item
Each user $k\in [K]$ estimates the demanded file as $\widehat{F}_k = \mu_{k}(X_{\mathbf{d}}, Z_{k})$, where 
\begin{align}
\mu_{k}:\ [0:1]^{\left\lceil RB\right\rceil }\times[0:1]^{\left\lfloor MB \right\rfloor} \rightarrow[0:1]^{B}, \ \forall k\in[K].
\label{eq:caching dec def}
\end{align}

\item
The (worst-case over all possible demands) probability of error is  
\begin{align}
P_\text{e}^{(B)}\vcentcolon=\max_{\mathbf{d}\in[N]^{K}}\Pr\left[ 
\bigcup_{k=1}^{K} \left\{\widehat{F}_k \neq F_{d_{k}}\right\} 
\right].
\label{eq:caching Pe cent def}
\end{align}


\item
A pair $(M,R)$ is said to be achievable
if there exit placement functions as in~\eqref{eq:caching placement def}, encoding function as in~\eqref{eq:caching enc def} and decoding functions as in~\eqref{eq:caching dec def} such that $\lim_{B\to\infty} P_\text{e}^{(B)}= 0$, where $P_\text{e}^{(B)}$ was defined in~\eqref{eq:caching Pe cent def}.

\item
The objective is to determine, for a fixed $M$, 
the (worst-case) load
\begin{align}
R^{\star} \vcentcolon=\inf\{R : (M,R) \ \text{is achievable} \}.
\label{eq:caching cent load def}
\end{align}
 
\end{itemize}
 
In the following, we say that the placement phase is \emph{uncoded} if the bits of the various files are simply copied within the caches. Formally,
\begin{defn}[Uncoded cache placement]
\label{def:uncoded cache placement}
The placement phase is said to be uncoded if the cache contents in~\eqref{eq:caching placement def} are  
$Z_{k} = (A_{1,k},A_{2,k},\ldots,A_{N,k})$ where $A_{i,k}\subseteq F_{i}$ for all files $i\in[N]$ and such that $\sum_{i\in[N]}|A_{i,k}| \leq MB$, for all users $k\in[K]$.
\end{defn}
The (worst-case) load under the constraint of uncoded cache placement 
is denoted as $R_{\textrm{u}}^{\star}$.
Trivially $R_{\textrm{u}}^{\star} \geq R^{\star}.$

\subsection{The Centralized Caching Problem: MAN Achievability}
\label{sec:model:cache in}

We start with the description of the MAN scheme in~\cite{dvbt2fundamental}.
Let the cache size be $M=t\frac{N}{K}$, for some positive integer $t\in[0:K]$. 
In the placement phase, each file is partitioned into $\binom{K}{t}$ 
equal size sub-files
of $B/\binom{K}{t}$ bits. The sub-files of $F_{i}$ are denoted by $F_{i,\Wc}$
for $\Wc\subseteq[K]$ where $|\Wc|=t$. 
User $k\in [K]$ 
fills its cache as
\begin{align}
Z_k 
= \Big( 
F_{i,\Wc}
:  k\in\Wc, 
\ \Wc\subseteq[K], 
\ |\Wc|=t,
\ i\in[N]
\Big).
\label{eq:cMAN cache function}
\end{align}
In the delivery phase, given the demand vector $\mathbf{d}$, the server transmits
\begin{align}
X_{\mathbf{d}} 
= \Big( \oplus_{s\in\Sc}F_{d_{s},\Sc\backslash\{s\}}
: \Sc\subseteq[K], |\Sc|=t+1 
\Big),
\label{eq:cMAN tx signal}
\end{align}
which requires broadcasting $B \binom{K}{t+1}/\binom{K}{t}$ bits.
Note that user $k\in\Sc$, for $\Sc$ as in~\eqref{eq:cMAN tx signal},
wants $F_{d_{k},\Sc\backslash\{k\}}$ and has cached $F_{d_{s},\Sc\backslash\{s\}}$ 
for all $s\in\Sc : s\neq k$, so it can recover $F_{d_{k},\Sc\backslash\{k\}}$ 
from $X_{\mathbf{d}}$ in~\eqref{eq:cMAN tx signal} and the cache content in~\eqref{eq:cMAN cache function}.
The load is thus given by 
\begin{align}
R_\textrm{c,u,MAN}[t] \vcentcolon=  
\frac{\binom{K}{t+1}}{\binom{K}{t}} \geq R^{\star},
\label{eq:cMANloadupper}
\end{align}
where the subscript ``{c,u,MAN}'' in~\eqref{eq:cMANloadupper} stands for ``centralized uncoded-placement Maddah-Ali and Niesen.'' For $M \frac{K}{N}$ not an integer, one takes the lower convex envelope of the set of points $(M,R)=\left(t \frac{N}{K}, R_\textrm{c,u,MAN}[t]\right)$  for $t \in[0:K]$.

The MAN scheme was improved in~\cite{yu2017characterizing} as follows.
Of the $\binom{K}{t+1}$ transmitted linear combinations in~\eqref{eq:cMAN tx signal}, $\binom{K-\min(K,N)}{t+1}$ can be obtained as linear combinations of the other transmissions. Therefore, by removing these redundant transmissions for the case $N<K$, the load becomes
\begin{align}
 R_\textrm{c,u,YMA}[t] \vcentcolon= 
\frac{ \binom{K}{t+1} - \binom{K-\min(K,N)}{t+1} }{ \binom{K}{t} } \geq R^{\star},
\label{eq:cYMAloadupper}
\end{align}
for $M = t \frac{N}{K}$ with $t\in[0:K],$ 
where the subscript ``{c,u,YMA}'' in~\eqref{eq:cYMAloadupper} stands for ``centralized uncoded-placement Yu Maddah-Ali Avestimehr.'' For $M \frac{K}{N}$ not an integer, one takes the lower convex envelope of set of points $(M,R)=\left(t \frac{N}{K}, R_\textrm{c,u,YMA}[t]\right)$ for $t \in[0:K]$. Notice that the load in~\eqref{eq:cYMAloadupper} is strictly smaller than the one in~\eqref{eq:cMANloadupper} for $N < K$. 

\subsection{The Index Coding Problem: Definition}
\label{sec:model:IC}

\begin{figure}
\centering{}
\includegraphics[scale=0.5]{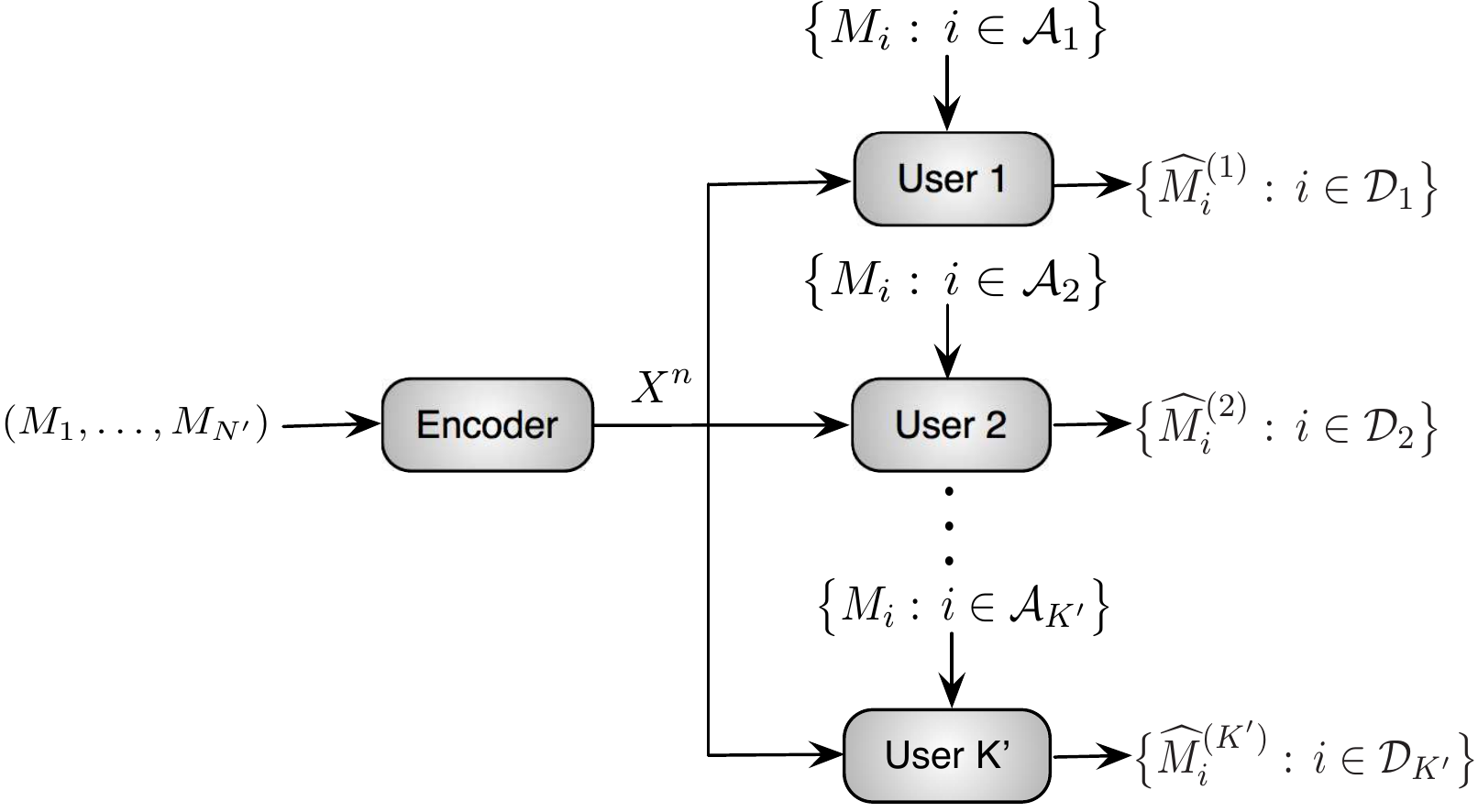}
\caption{\small 
An IC problem with $N^{\prime}$ files and $K^{\prime}$ users.
}
\label{fig:index coding}
\end{figure}

The IC problem, shown in Fig.~\ref{fig:index coding} and originally proposed in~\cite{birk1998informedsource} in the context of broadcasting with message side information, is defined as follows.
\begin{itemize}

\item
A sender wishes to communicate $N^{\prime}$ independent messages to $K^{\prime}$ users. 
The server is connected to the users through a noiseless channel with finite input alphabet $\mathcal{X}$. 

\item
Each user $j\in [K^{\prime}]$ demands a set of messages indexed by $\Dc_{j} \subseteq [N^{\prime}]$ and knows a set of messages indexed by $\Ac_{j} \subseteq [N^{\prime}]$. In order to avoid trivial problems, it is assumed that $\Dc_{j} \not= \emptyset$, $\Ac_{j} \not= [N^{\prime}]$, and $\Dc_{j}\cap\Ac_{j}=\emptyset$. 

\item
A $(|\mathcal{X}|^{nR_{1}},\ldots,|\mathcal{X}|^{nR_{N^{\prime}}},n,\epsilon_n)$-code 
for the IC problem is defined as follows.
Each message $M_{i}$, $i\in [N^{\prime}],$ is uniformly distributed on  $[|\mathcal{X}|^{nR_{i}}]$,
where $n$ is the block-length and $R_{i} \geq 0$ is the transmission rate in symbols per channel use.
In order to satisfy the users' demands, the server broadcasts $X^{n}=\mathsf{enc}(M_{1},\ldots,M_{N^{\prime}})\in \mathcal{X}^{n}$ where $\mathsf{enc}$ is the encoding function.
Each user $j\in[K^{\prime}]$ estimates the messages indexed by $\Dc_{j}$ by the decoding function $\mathsf{dec}_j\big( X^{n}, (M_{i}:i\in\Ac_{j}) \big)$. The probability of error is
\begin{align}
\epsilon_n \vcentcolon
&= \max_{j\in[K^{\prime}]}\Pr\left[\mathsf{dec}_j\big( X^{n}, (M_{i}:i\in\Ac_{j}) \big) \not= 
\right. \nonumber\\ &\left. 
(M_{i}:i\in\Dc_{j}) \right]. 
\label{eq:errIC}
\end{align}

\item
A rate vector $(R_{1},\ldots,R_{N^{\prime}})$ is said to be achievable if there exists a family of $(|\mathcal{X}|^{nR_{1}},\ldots,|\mathcal{X}|^{nR_{N^{\prime}}},n,\epsilon_n)$-codes 
for which $\lim_{n\to\infty}\epsilon_n=0$, for $\epsilon_n$ in~\eqref{eq:errIC}.

\item
The goal is to find the capacity region, defined as the largest possible set of achievable rate vectors.

\end{itemize}

\begin{rem}\label{rem:|X| does not matter}
We used the definitions in~\cite[Chapter 1, Section 1.2]{fundamentalindexcoding}, in which the definition of capacity region depends on the alphabet size $|\mathcal{X}|$. However, as proved in \cite[Lemma~1.1]{fundamentalindexcoding}, the choice of the alphabet size $\mathcal{X}$ is irrelevant to the actual capacity region itself. Intuitively, this is so because the rates are defines in symbols per channel use or, equivalently, the base of the logarithms is $|\mathcal{X}|$. 
\end{rem}

\subsection{The Index Coding Problem: Composite (Index) Coding Achievable Bound}
\label{sec:model:IC in}
The composite (index) coding achievable bound proposed in~\cite{onthecapacityindex} is a two-stage scheme based on binning and non-unique decoding. In the first encoding stage, for each $\Jc\subseteq[N^{\prime}]$, the messages $(M_{i} : i\in\Jc)$ are encoded into the \emph{composite index} $W_{\Jc}\in[|\Xc|^{nS_{\Jc}}]$ at some rate $S_{\Jc} \geq 0$  based on random binning. By convention, $S_{\emptyset}=0$. 
In the second encoding stage, 
the collection of all composite indices $(W_{\Jc} : \Jc\subseteq[N^{\prime}])$ is mapped into a length-$n$ 
sequence $X^{n}$ which is received error-free by all users. 

In the first decoding stage, every user recovers all composite indices. 
In the second decoding stage, user $j\in[K^{\prime}]$ chooses a set $\Kc_{j}$ such that $\Dc_{j} \subseteq  \Kc_{j} \subseteq [N^{\prime}]\backslash\Ac_{j}$ 
and uniquely decodes all messages $(M_{i} : i\in\Kc_{j})$;
the decoding of user $j\in[K^{\prime}]$ is based on the recovered $(W_{\Jc} : \Jc\subseteq \Kc_{j}\cup\Ac_{j})$.

The achievable rate region by composite (index) coding is stated next, from~\cite[Proposition 6.11]{onthecapacityindex}.
\begin{thm}[Composite IC Achievable Bound, generalized to allow for multicast messages]
\label{thm: thm2 composite coding}
\begin{subequations} 
A non-negative rate tuple $\mathbf{R} \vcentcolon=(R_{1},\ldots,R_{N^{\prime}})$ is achievable
for the IC problem $\big( (\Ac_{j},\Dc_{j}) : j\in[K^{\prime}] \big)$ defined in Section~\ref{sec:model:IC} provided that
\begin{align}
&\mathbf{R}\in
\bigcap_{j\in [K^{\prime}]} \quad
\bigcup_{\small
\Kc_{j}:
\Dc_{j}\subseteq \Kc_{j}\subseteq [N^{\prime}]\backslash \Ac_{j}
}
\mathscr{R}(\Kc_{j}|\Ac_{j},\Dc_{j}),
\label{eq:composite 1}
\\
&\mathscr{R}(\Kc|\Ac,\Dc)
\vcentcolon=  \bigcap_{\Jc : \Jc\subseteq\Kc } 
\left\{ \sum_{i\in\Jc}R_{i} < v_{\Jc} \right\},
\label{eq:composite 2}
\end{align}
where in~\eqref{eq:composite 2} $v_{\Jc}$ is defined as
\begin{align}
v_{\Jc}\vcentcolon=\sum_{\Pc: \Pc\subseteq\Ac\cup\Kc , \Pc\cap\Jc\neq\emptyset} 
S_{\Pc}, %
\label{eq:composite 3 vJ}
\end{align} 
and where in~\eqref{eq:composite 3 vJ} the non-negative quantities $(S_{\Jc}:\Jc\subseteq[N^{\prime}])$ must satisfy 
\begin{align}
\sum_{\Jc:\Jc\in[N^{\prime}], \Jc\nsubseteq\Ac_{j}}S_{\Jc} < 1, \quad \forall j\in [K^{\prime}].
\label{eq:composite 4 decompression}
\end{align}
\label{eq:thm2 composite coding}
\end{subequations} 
\end{thm}
Note that the constrain in~\eqref{eq:composite 4 decompression} is from the first decoding stage 
and the one in~\eqref{eq:composite 3 vJ} from the second decoding stage. 

\subsection{The Index Coding Problem: Acyclic Subgraph Converse Bound for Multiple Unicast Index Coding}
\label{sec:model:IC out}

If $K^{\prime} = N^{\prime}$ and 
$\Dc_{j}=\{j\}$ where $j\not\in \Ac_{j}$ for each $j\in[N^{\prime}]$, 
the IC is known as the \emph{multiple unicast IC}.
The multiple unicast IC problem 
can be represented as a directed graph $G$, where each node in the graph represents one user and its demanded message,
and where a directed edge connects node~$i$ to node~$j$ if user~$j$ knows the message desired by user~$i$. 
By the 
submodularity of entropy, 
a converse bound was proposed in~\cite[Theorem 3.1]{Lexicographic} for the symmetric rate case 
and extended 
in~\cite[Theorem 1]{onthecapacityindex} to the case where messages can have different rates.
Due to the high computational complexity, the converse bound in~\cite[Theorem 1]{onthecapacityindex} can only be evaluated for IC problems with limited number of messages. 
A looser (compared to~\cite[Theorem 1]{onthecapacityindex}) converse bound was proposed in~\cite[Corollary 1]{onthecapacityindex} and is stated next.

\begin{thm}[Acyclic Subgraph Converse Bound for Multiple Unicast IC~\cite{onthecapacityindex}]
\label{thm:uncycle bound} 
If $(R_{1},\ldots,R_{N^{\prime}})$ is achievable for the multiple unicast IC problem $\big( (\mathcal{A}_{j},\mathcal{D}_{j}=\{j\}):j\in [N^{\prime}]\big)$, defined in Section~\ref{sec:model:IC} and represented by the directed graph $G$, then it must satisfy
\begin{align}
\sum_{j\in\mathcal{J}} R_{j} \leq 1,
\label{eq:Rj<1}
\end{align}
for all $\mathcal{J}\subseteq[N^{\prime}]$ where the sub-graph of $G$ over the vertices in
$\mathcal{J}$ does not contain a directed cycle.
\end{thm}  

The proof of Theorem~\ref{thm:uncycle bound} is based on noticing that a user~$k_1$ in the found acyclic subgraph can decode the message of the following user~$k_2$ either because $\mathcal{A}_{k_2} \subseteq \mathcal{D}_{k_1}\cup\{k_1\}$ (user~$k_1$ can mimic user~$k_2$) or by giving user~$k_1$ genie side information so that it can mimic user~$k_2$.

\subsection{Mapping the Caching Problem with Uncoded Cache Placement into an Index Coding Problem}
\label{sec:model:connection}
As mentioned before, the caching problem with uncoded cache placement can be see as a family of IC problems.
The difference between caching and IC is that the side information sets are fixed in IC, while they represent the cache contents that must be properly designed in caching; moreover, in IC the demands are also fixed, while in caching one must consider all possible demands. In caching, if the cache placement phase is uncoded, the delivery phase is an IC problem for appropriately defined message and demand sets. Hence, IC results can be leveraged for the caching problem, as we do in this paper.

Under the constraint of uncoded cache placement, when the cache contents and the demands are fixed, 
the delivery phase of the caching problem is equivalent to the following IC problem. 
Denote the set of distinct demanded files in the demand vector $\mathbf{d}$ by $\mathcal{N}(\mathbf{d})$,
that is $\mathcal{N}(\mathbf{d}) := \cup_{k\in[K]}\{d_k\}$.
For each $i\in\mathcal{N}(\mathbf{d})$ and for each $\Wc\subseteq[K]$, 
the sub-file $F_{i,\Wc}$ (containing the bits of file $F_{i}$ only within the cache of the users indexed by $\Wc$)
is an independent message in the IC problem with user set $[K]$.
Hence, by using the notation introduced in Sections~\ref{sec:model:IC} and~\ref{sec:model:caching},
we have 
$K^{\prime} = K$ and 
$N^{\prime} \leq |\mathcal{N}(\mathbf{d})| (2^K-1)$. 
For each user $k\in [K]$ in this general IC problem, 
the desired message set and the side information sets are given by   
\begin{align}
\Dc_{k}&\vcentcolon= \big\{ F_{d_{k},\Wc}: \Wc\subseteq[K], k\notin\Wc \big\}, 
\label{eq:dk def for caching}
\\
\Ac_{k}&\vcentcolon= \big\{ F_{i,\Wc} : \Wc\subseteq[K], i\in\mathcal{N}(\mathbf{d}), k\in \Wc \big\}.
\label{eq:ak def for caching}
\end{align}

\section{Converse Bound for Centralized Cache-aided Systems with Uncoded Cache Placement}
\label{sec:centralized converse}

In this section, we leverage the connection between caching and IC problems outlined in Section~\ref{sec:model:connection} to investigate the fundamental limits of centralized cache-aided systems with uncoded cache placement. We derive a converse bound (by using Theorem~\ref{thm:uncycle bound} in Section~\ref{sec:model:IC out}) that matches the achievable load $R_\textrm{c,u,YMA}$ in~\eqref{eq:cYMAloadupper}.

\subsection{Theorem Statement}
\label{sec:centralized:outer bound}

The following converse bound on the load of centralized cache-aided systems under the constraint of uncoded cache placement was first presented in our conference papers~\cite{ontheoptimality,ourisitinnerbound}. 
Recall that we denote the optimal load by $R^{\star}$, 
and the optimal load under the constraint of uncoded cache placement (see Definition~\ref{def:uncoded cache placement}) as $R_{\textrm{c,u}}^{\star}$. 
\begin{thm}
\label{thm:centralized outer bound} 
In centralized cache-aided systems 
the load $R_{\textrm{c,u}}^{\star}$ satisfies 
\begin{align}
R_{\textrm{c,u}}^{\star} 
&\geq c_{q} + (c_{q}-c_{q-1}) \left(\frac{KM}{N}-q\right),
\label{eq:Rcen outer bound}
\\
c_{q} &\vcentcolon= \frac{\binom{K}{q+1}-\binom{K-\min(K,N)}{q+1}}{\binom{K}{q}},
  \quad \forall q\in[0:K].
\label{eq:cq in outer bound}
\end{align}
Moreover, this converse bound 
is a piece-wise linear curve with corner points
\begin{align}
(M,R)=\left(q\frac{N}{K},
c_{q} \right), \quad \forall q\in[0:K].
\label{eq:corners in outer bound}
\end{align}
\end{thm}
%

Before we proceed to prove the general converse bound in Theorem~\ref{thm:centralized outer bound}, we give a specific example. This example introduces the main ideas in the proof.

\subsection{An Example}

\begin{figure*}
\centering{}
\includegraphics[scale=0.35]{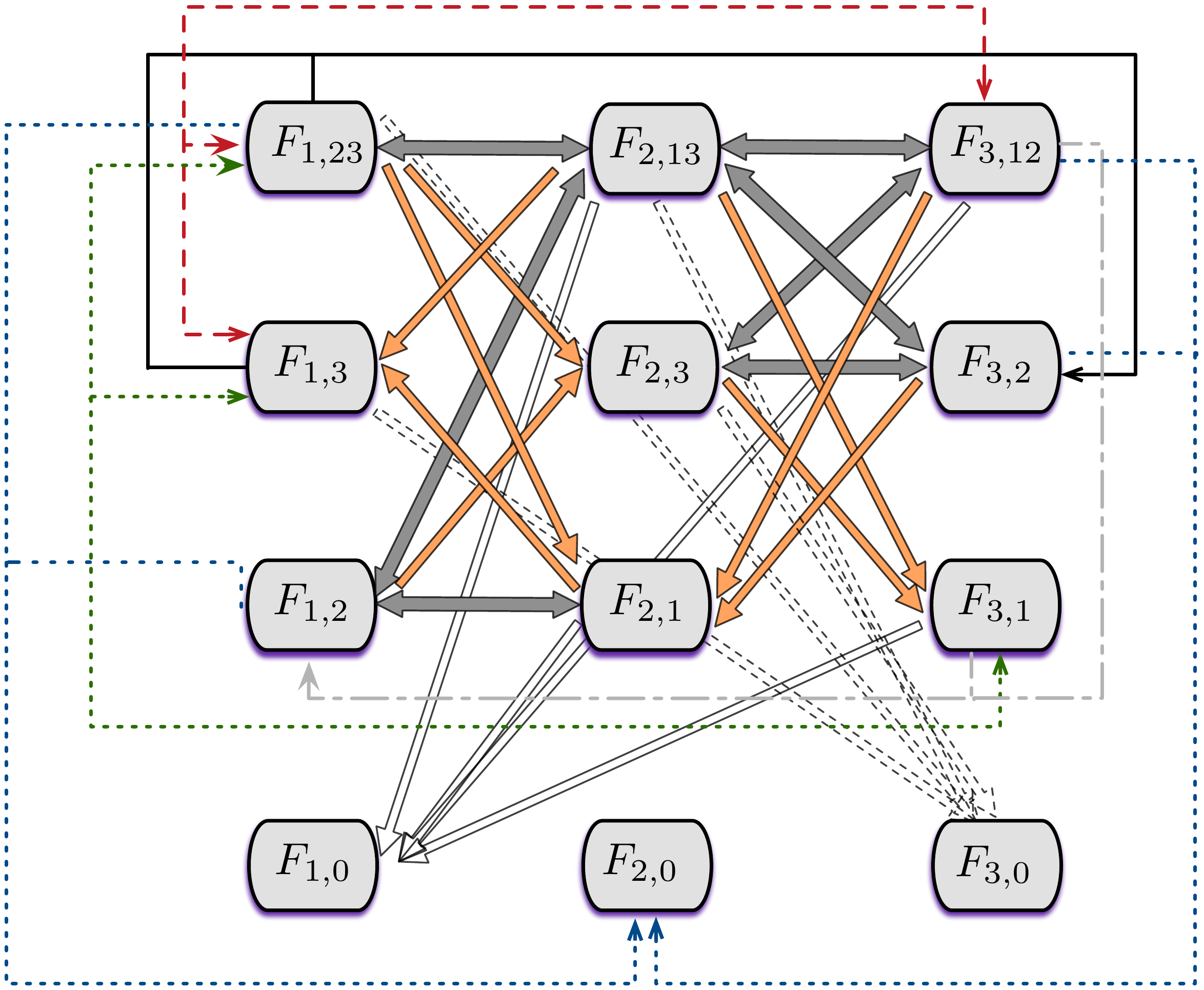}
\caption{\small Directed graph for the equivalent IC scenario corresponding to the caching problem with $N=K=3$, and with demand vector $\mathbf{d}=(1,2,3)$. Each sub-file demanded by each user is an independent node in the directed graph. A directed arrow from $F_{d_{k_1},\Wc_{1}}$ to $F_{d_{k_2},\Wc_{2}}$ appears if and only if user $k_2$ caches $F_{d_{k_1},\Wc_{1}}$ (i.e., $k_2\in \Wc_{1}$). Different types and colours for arrows are intended to distinguish dense arrows in the figure.}\vspace{2mm}
\label{fig:case3graph}
\includegraphics[scale=0.3]{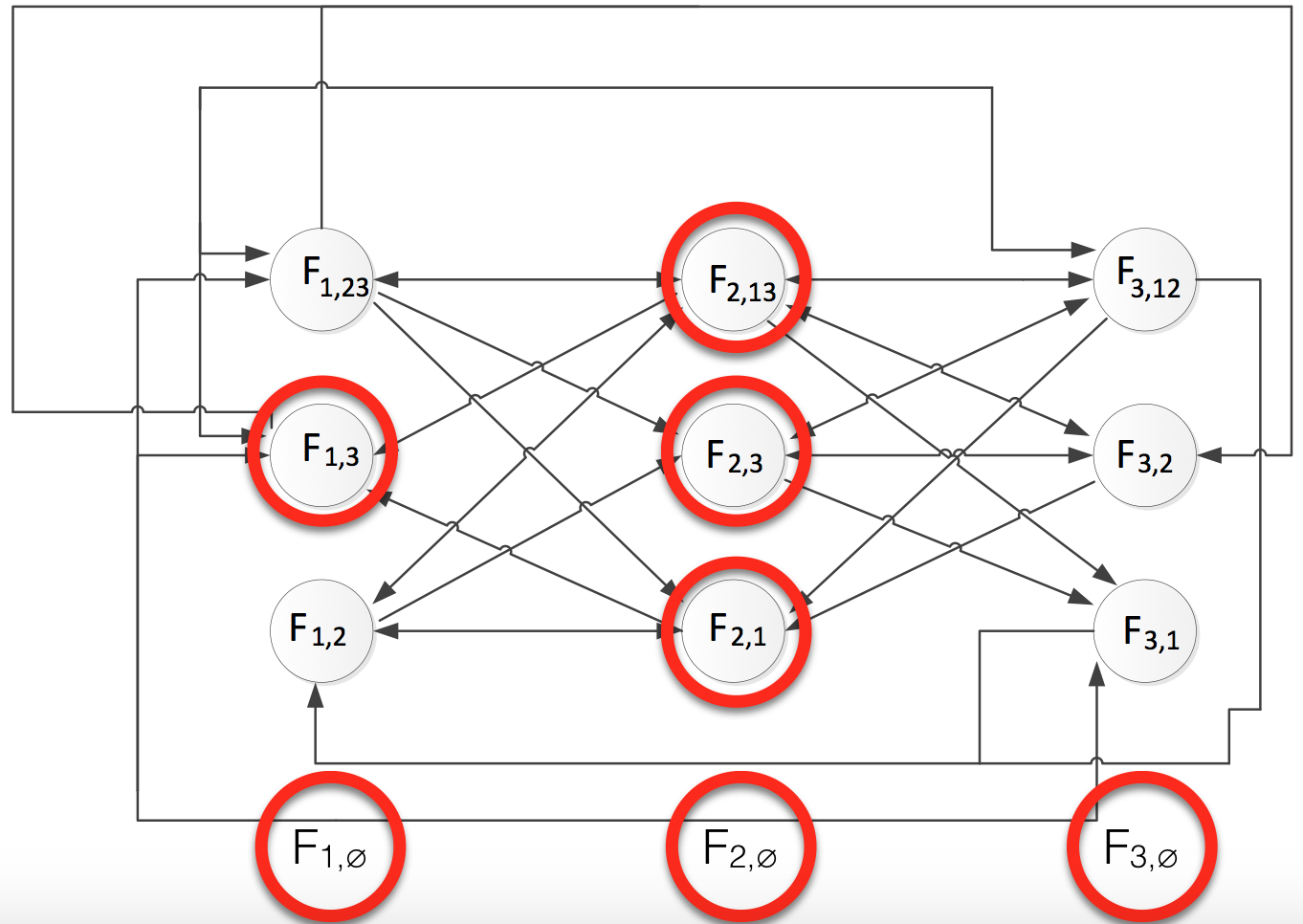}
\includegraphics[scale=0.3]{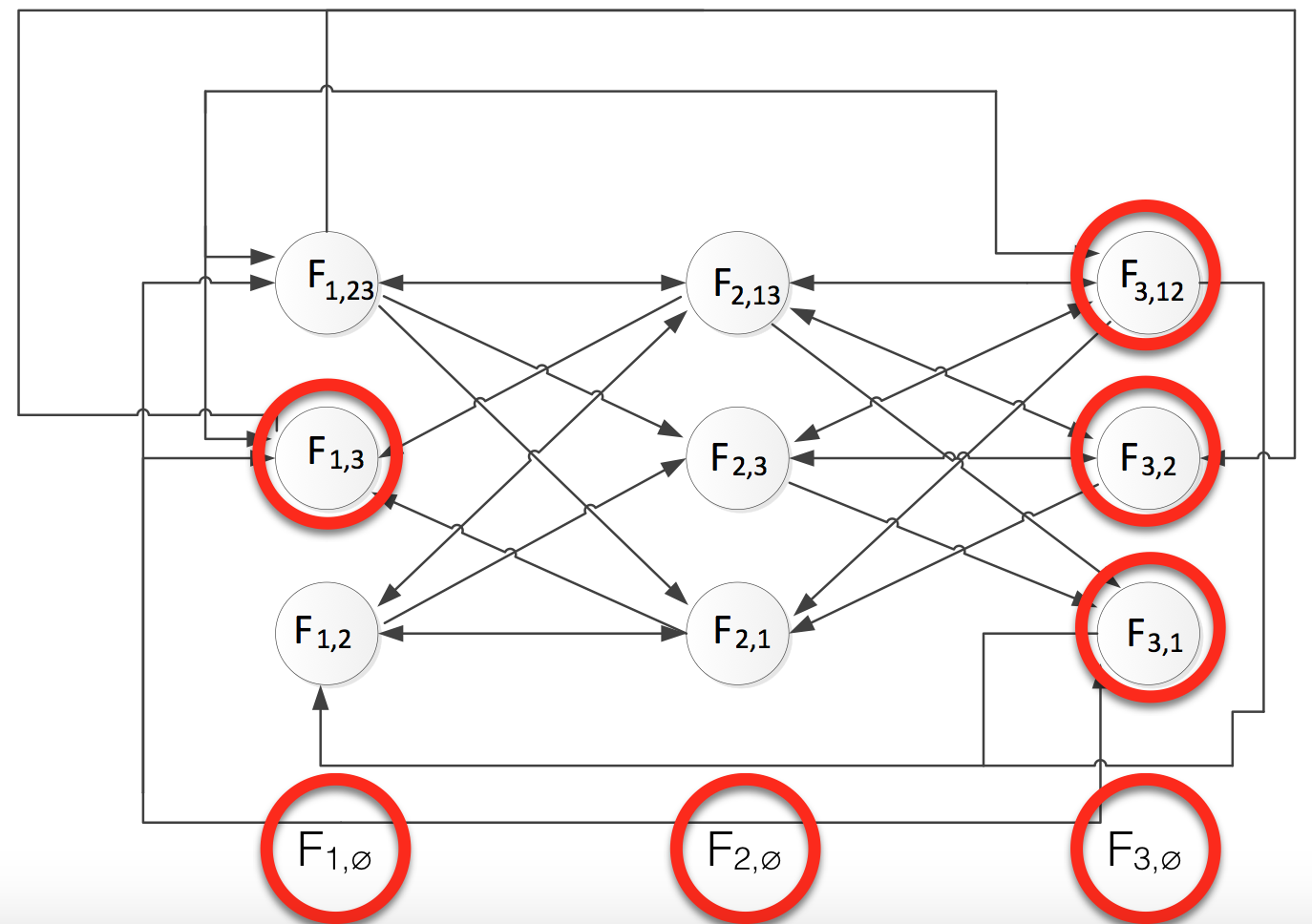}
\caption{\small  Nodes forming an acyclic subgraph are circled in red.
Left:  demand $(1,2,3)$ and permutation $(2,1,3)$.
Right: demand $(1,2,3)$ and permutation $(3,1,2)$.}
\label{fig:case3graphperms}
\end{figure*}

The reasoning in this example applies to the general case $K\leq N$;
the case $K> N$ will be dealt with in the general proof.

Assume that the server has $N=K=3$ files, denoted as $(F_{1},F_{2},F_{3})$.
The file length in number of bits is $B$.
The cache size in number of bits is $MB$,
for some $M\in[0,N]=[0,3]$.
After the uncoded cache placement phase is done, each file $F_{i}$ can be thought as having been divided into $2^{K}=2^3=8$ disjoint sub-files, 
denoted as 
\begin{align}
\big( F_{i,\mathcal{W}} : \mathcal{W}\subseteq[K], \ i\in[N] \big),
\end{align}
where
$F_{i,\mathcal{W}}$ has been cached only by the users indexed by $\mathcal{W}$.
For simplicity in the following we omit the braces when we indicate sets, i.e., $F_{1,12}$ represents $F_{1,{\{1,2\}}}$, which dos not create any confusion for this example.

For each demand vector $\mathbf{d}=(d_{1},d_{2},d_{3})\in[N]^K=[3]^3$,
we generate an IC problem with at most $|\mathcal{N}(\mathbf{d})| 2^{K-1} =  12$ independent messages.
These messages are
\begin{align}
\bigcup_{k\in [K], \mathcal{W} \subseteq [K] : k\not\in \mathcal{W}} F_{ d_{k},\mathcal{W}},
\end{align}
and represents the sub-files demanded by the users in 
$[K]$ 
but not available in their caches.
The messages available as side information to user $k\in[K]$ for this IC are 
\begin{align}
\bigcup_{i\in\mathcal{N}(\mathbf{d}), \mathcal{W} \subseteq [K] : k\in \mathcal{W}} F_{i,\mathcal{W}}.
\end{align}
For this IC problem, we generate a directed graph as follows. 
Each vertex corresponds to a different sub-file. There is a directed arrow from $F_{d_{k_1},\Wc_{1}}$ to $F_{d_{k_2},\Wc_{2}}$ if and only if user $k_2$ caches $F_{d_{k_1},\Wc_{1}}$ (i.e., $k_2\in \Wc_{1}$).
For example, Fig.~\ref{fig:case3graph} shows the directed graph representing this IC problem for the demand vector $\mathbf{d}=(1,2,3)$. 

Consider the demand vector $\mathbf{d}=(d_{1},d_{2},d_{3})$, where $d_{i}\in[N]=[3], \ i\in[K]=[3]$.
In order to apply Theorem~\ref{thm:uncycle bound}, in the constructed directed graph we want to find sets of vertices $\mathcal{J}$ that do not form a directed cycle. 
No receiver has stored $F_{1,\emptyset},F_{2,\emptyset},F_{3,\emptyset}$, so there is no outgoing edge from $F_{1,\emptyset},F_{2,\emptyset},F_{3,\emptyset}$ to any other vertex in the graph. 
Therefore, $F_{1,\emptyset},F_{2,\emptyset},F_{3,\emptyset}$ are always in the such sets $\mathcal{J}$ when we evaluate~\eqref{eq:Rj<1}. 
%

We focus next on demand vectors $\mathbf{d}$ with distinct demands, that is, $|\mathcal{N}(\mathbf{d})|=\min(N,K)=K=3$;
the worst case demand may not be in such a set of demand vectors,
but this is not a problem as we aim to derive a converse bound on the (worst-case) load at this point.
For a demand vector $\mathbf{d}$ with distinct demands,
consider now a permutation $\mathbf{u}=(u_{1},u_{2},u_{3})$ of $[K]=[3]$.
For each such  $\mathbf{u}$, a set of nodes not containing a cycle is as follows: 
$F_{d_{u_{1}},\mathcal{W}_{1}}$ for all $\mathcal{W}_{1}\subseteq[K]\backslash \{u_{1}\}$, and
$F_{d_{u_{2}},\mathcal{W}_{2}}$ for all $\mathcal{W}_{2}\subseteq[K]\backslash \{u_{1},u_{2}\}$, and 
$F_{d_{u_{3}},\mathcal{W}_{3}}$ for all $\mathcal{W}_{3}\subseteq[K]\backslash \{u_{1},u_{2},u_{3}\}=\emptyset$. 
 For example, when $\mathbf{d}=(1,2,3)$ and $\mathbf{u}=(1,3,2)$, we have
\begin{align}
&d_{u_1}=d_{1}=1; \mathcal{W}_1\subseteq[K]\backslash \{u_1\}=[3]\backslash \{1\}=\{2,3\},
\\
&
d_{u_2}=d_{3}=3; \mathcal{W}_2\subseteq[K]\backslash \{u_1,u_2\}=[3]\backslash \{1,3\}=\{2\},
\\
&
d_{u_3}=d_{2}=2; \mathcal{W}_3\subseteq[K]\backslash \{u_1,u_2,u_3\}
=\emptyset,
\end{align}
and the corresponding set not containing a cycle is 
\begin{align}
(F_{1,\emptyset},F_{1,2},F_{1,3},F_{1,23},F_{3,\emptyset},F_{3,2},F_{2,\emptyset}).
\label{eq:acex}
\end{align} 
More precisely, $F_{1,\emptyset},F_{1,2},F_{1,3},F_{1,23}$ are demanded by user $1$ and thus there is no cycle among them. User $1$ does not know $F_{3,\emptyset},F_{3,2}$ and thus there is no directed arrow from  each of $F_{3,\emptyset},F_{3,2}$ to each of $F_{1,\emptyset},F_{1,2},F_{1,3},F_{1,23}$. So there is no cycle in $(F_{1,\emptyset},F_{1,2},F_{1,3},F_{1,23},F_{3,\emptyset},F_{3,2})$. Finally, user $1$ and user $3$ do not know $F_{2,\emptyset}$, and thus there is no directed arrow from $F_{2,\emptyset}$ to each of $F_{1,\emptyset},F_{1,2},F_{1,3},F_{1,23},F_{3,\emptyset},F_{3,2}$. So we can see the chosen set is acyclic (see figure~\ref{fig:case3graphperms}).

From~\eqref{eq:Rj<1}, we have that the acyclic set of nodes in~\eqref{eq:acex} can be used to write the following bound 
(in which $2^{B R_\textrm{c,u}^{\star}}$ plays the role of $|\mathcal{X}|$)
\begin{align}
&B R_\textrm{c,u}^{\star} \geq \nonumber\\
&|F_{1,\emptyset}|+|F_{1,2}|+|F_{1,3}|+|F_{1,23}|+|F_{3,\emptyset}|+|F_{3,2}|+|F_{2,\emptyset}|.
\label{eq:example of acyclic bound}
\end{align}
In general, when $K\leq N$ we can find a bound such as the one in~\eqref{eq:example of acyclic bound} for all possible pairs $\mathbf{d}\in\textrm{Perm}(N,K)$ and $\mathbf{u}\in \textrm{Perm}(K,K)$,
where $\textrm{Perm}(n,k)$ denotes the set of all $k-$permutations of $n$ elements (there are $\frac{n!}{(n-k)!}$ elements in the set $\textrm{Perm}(n,k)$  for $n\geq k$).
%
\begin{subequations}
If we sum all the $|\textrm{Perm}(N,K)| |\textrm{Perm}(K,K)| =  \binom{N}{K} (K!)^2 =(3!)^2 = 36$ inequalities as in~\eqref{eq:example of acyclic bound}, we get
\begin{align}
  R_\textrm{c,u}^{\star} &\geq \frac{1}{(3!)^2 }
\sum_{\mathbf{d} \in \textrm{Perm}(3,3)} 
\sum_{\mathbf{u} \in \textrm{Perm}(3,3)} 
\sum_{j \in [3] } 
\sum_{\mathcal{W}_j \subseteq [3] \backslash \{u_1,\dots,u_j\}} 
\!\!\!\!\!\!\!\!
\frac{|F_{d_{u_{j}}, \mathcal{W}_j}|}{B}
\label{eq:case3inequal step1}
\\&=
\sum_{t \in [0:3]} x_{t} \frac{\binom{3}{t+1}}{\binom{3}{t}} 
\label{eq:case3inequal step2}
\\&= 3 \cdot x_{0}+ 1 \cdot x_{1}+\frac{1}{3} \cdot x_{2}+0 \cdot x_{3},
\label{eq:case3inequal step3}
\end{align}
\label{eq:case3inequal}
\end{subequations}
where $x_{t}$ in~\eqref{eq:case3inequal step2} is defined as
\begin{align}
0\leq x_{t}\vcentcolon=\sum_{j\in[N]} 
\ \sum_{\mathcal{W}\subseteq [K] : |\mathcal{W}|=t}  
\frac{|F_{j,\mathcal{W}}|}{NB}, \quad t \in [0:K],
\label{eq:defxi}
\end{align}
and represnts the fraction of the total 
number of bits across all files
that are known/cached exclusively by a subset of $t\in[0:K]=[0:3]$ users.


The general proof of~\eqref{eq:case3inequal step2} can be found in~\eqref{eq:step 2 final ineq}.
At this point we can offer the following intuitive interpretation for the case $K\leq N$, as it is the case in this example.
The total number of sub-files cached by a subset of $t\in[0:K]$ users is $N\binom{K}{t}$, 
where the factor $\binom{K}{t}$ appears at the denominator of~\eqref{eq:case3inequal step2} (here $K=3$), and
the factor $N$ at the denominator of~\eqref{eq:defxi}.
The total number of sub-files cached by a subset of $t\in[0:K]$ users 
 in~\eqref{eq:example of acyclic bound} (by the symmetry of the problem,   the other bounds for different pair $\mathbf{d}$ and $\mathbf{u}$ have the same structure as~\eqref{eq:example of acyclic bound}) is 
\begin{align}
\sum_{i\in[K]} \sum_{\Wc_{i}\subseteq[K]\backslash\{u_{1},\ldots,u_{i}\}} 1_{\{|\Wc_{i}| = t\}}
\notag\\=
\binom{K-1}{t} + \binom{K-2}{t} + \ldots +\binom{t}{t} = \binom{K}{t+1},
\end{align}
where $1_{\{\Ac\}}$ in the indicator function that is equal to one if and only if the condition in $\Ac$ is true,
and where we use the Pascal's triangle identity;
the factor $\binom{K}{t+1}$ (here $K=3$) appears at the numerator of~\eqref{eq:case3inequal step2}.

In addition to the bounds in~\eqref{eq:case3inequal} and~\eqref{eq:defxi},
we also have that the total number of bits in the files is
\begin{align}
\sum_{j\in[N]} \sum_{\mathcal{W}\subseteq[K]}  |F_{j,\mathcal{W}}| = NB  
\Longleftrightarrow
\sum_{t\in[K]} x_{t} = 1,
\label{eq:casefilesize}
\end{align}
and that the total number of bits in the caches must satisfy
\begin{align}
\sum_{j\in[N]} \sum_{\mathcal{W}\subseteq[K] : j\in \mathcal{W}}  |F_{j,\mathcal{W}}| \leq KMB
\Longleftrightarrow
\sum_{t\in[K]} t \ x_{t} \leq \frac{KM}{N}.
\label{eq:case3cachesize}
\end{align}

Please note that~\eqref{eq:casefilesize} arises from the total number of bits across all files,
which is a looser constraint than imposing that each file contains the same number of bits;
similarly,~\eqref{eq:case3cachesize} arises from the total number of bits across all caches,
which is a looser constraint than imposing that each cache has the same size. 
None of these is a problem as we aim to derive a converse bound on the load at this point.
This implies that that the converse bound we derive applies to the case where the total number of bits across all files and the total number of bits across all caches are constrained, but not the size each individual file or each individual cache.

The constraints in~\eqref{eq:case3inequal}-\eqref{eq:case3cachesize}
provide a converse bound for the load $R_\textrm{c,u}^{\star}$ with uncoded cache placement. 
Since there are many inequalities in $K+1=4$ unknowns, we proceed to 
eliminate $(x_0,x_1,x_2,x_3)$ in the system of inequalities in~\eqref{eq:case3inequal}-\eqref{eq:case3cachesize}. 
By doing so, we obtain
\begin{align}
 R_\textrm{c,u}^{\star} 
 &\stackrel{\textrm{by eq.\eqref{eq:case3inequal step3}}}{\geq} 
  3x_{0}+  x_{1}+\frac{1}{3} x_{2}
\nonumber\\&
   \stackrel{\textrm{by eq.\eqref{eq:casefilesize}}}{=}
    3(1-x_1-x_2-x_3)+  x_{1}+\frac{1}{3} x_{2}
\nonumber\\&
   = 3 -  2x_{1}-\frac{8}{3} x_{2}-3x_3
\nonumber\\&
  \stackrel{\textrm{by eq.\eqref{eq:case3cachesize}}}{\geq}
   3 + 2(2x_2+3x_3-M)-\frac{8}{3} x_{2}-3x_3
\nonumber\\&=
   3-2M + \frac{2}{3} x_2 + 3x_3
\nonumber\\&
   \stackrel{\textrm{by eq.\eqref{eq:defxi}}}{\geq}
   3-2M.
 \label{eq:case3eq1}
\end{align}
Similarly, we can obtain
\begin{align}
R_\textrm{c,u}^{\star} &\geq  -\frac{2}{3}M+\frac{5}{3},
 \label{eq:case3eq2}
 \\
R_\textrm{c,u}^{\star}  &\geq  -\frac{1}{3}M+1.
 \label{eq:case3eq3}
\end{align}

The maximum among the right-hand sides of~\eqref{eq:case3eq1},
~\eqref{eq:case3eq2}
and~\eqref{eq:case3eq3}
give a piecewise linear curve with corner points:  $(0,3),(1,1),(2,\frac{1}{3}),(3,0)$.
Since these corner points are achieved by $R_\textrm{c,u,MAN}[t], t\in[0:3]$, 
in~\eqref{eq:cMANloadupper},
we conclude that the two-phase strategy in \cite{dvbt2fundamental} is optimal under the constraint of uncoded cache placement in this case. 
Note that this shows that demand vectors with distinct demands lead to the worst case load. 

We are now ready to extend the reasoning in this example to a general setting, where we do not necessarily impose $K\leq N$.

\subsection{Proof of Theorem~\ref{thm:centralized outer bound}}
\label{sub:converse proof}
Consider a system with uncoded cache placement  and a demand vector with $\min(K,N)$ distinct demanded files.  
We treat the delivery phase of this caching scheme as an IC problem, as described in Section~\ref{sec:model:connection}. 
We derive a converse bound on $R_{\textrm{c,u}}^{\star}$ 
using Theorem~\ref{thm:uncycle bound}. 
A directed graph can be generated for such IC problem as described in Section~\ref{sec:model:connection}.
We propose the following lemma to give the sets of nodes not containing a directed cycle, whose proof is in Appendix~\ref{sec:proof of lemma1}. 
\begin{lem}
\label{lem:For-each-graph}
 Let $\mathbf{u}=(u_{1},u_{2},\dots,u_{\min(K,N)})$ be a permutation of $\mathcal{C}$, where $\mathcal{C}$ is the chosen user set with different demands.
A set of nodes not containing a directed cycle in the directed graph of the corresponding
IC problem can be composed of sub-files
\begin{align}
\big(F_{d_{u_{i}},\Wc_{i}} : 
 \Wc_{i}\subseteq[K]\backslash \{u_{1},\ldots,u_{i}\},
 \ i\in[\min(K,N)]
 \big).
 \label{eq:lemma-acyclic-set}
\end{align} 
\end{lem}

With the set of nodes not containing a directed cycle in the directed graph of the corresponding
IC problem as in~\eqref{eq:lemma-acyclic-set}, we write the bound in~\eqref{eq:Rj<1} from Theorem~\ref{thm:uncycle bound} as
\begin{align}
R_{\textrm{c,u}}^{\star} 
&\geq \sum_{i\in[\min(K,N)]} \sum_{\Wc_{i}\subseteq[K]\backslash\{u_{1},\ldots,u_{i}\}}   \frac{ |F_{d_{u_{i}},\Wc_{i}}| }{B}.
\label{eq:original uncycle}
\end{align}
  Note that, in the bound in~\eqref{eq:original uncycle},  there are $\sum_{i\in[\min(K,N)]} \binom{K-i}{t}$ subfiles known by exactly $t$ users whose coefficient is $1$. We can also note that  in general there are $N\binom{K}{t}$ subfiles known by exactly $t$ users. 
\begin{subequations}
By considering all sets $\mathcal{C}$ of users with $\min(K,N)$ 
distinct demands, and all the permutations $\mathbf{u}$ of $\mathcal{C}$, 
we can list all the inequalities in the form of~\eqref{eq:original uncycle} and sum them together to obtain 
\begin{align}
R_{\textrm{c,u}}^{\star} 
&\geq \sum_{t\in[0:K]}\frac{\binom{K-1}{t}+\binom{K-2}{t}+\cdots+\binom{K-\min(K,N)}{t}}{\binom{K}{t}}
\ x_{t}
\label{eq:step 1 final ineq}
\\
& = \sum_{t\in[0:K]}\frac{\binom{K}{t+1}-\binom{K-\min(K,N)}{t+1}}{\binom{K}{t}}
\ x_{t},
\label{eq:step 2 final ineq}
\end{align}
where from~\eqref{eq:step 1 final ineq} to~\eqref{eq:step 2 final ineq} we use the Pascal's triangle equality,
where the set of coefficients $(x_0, \ldots, x_K)$ defined in~\eqref{eq:defxi} can be interpreted as a probability mass function (see~\eqref{eq:casefilesize}) subject to a first-moment constraint (as given in~\eqref{eq:case3cachesize}).
\end{subequations}

Next, we introduce the following key Lemma, whose proof can be found in Appendix~\ref{sec:proof of lemma2}.
\begin{lem}
\label{lem:monotony of Y}
Let $K$, $N$ be positive integers where $K>N$.
For any $q\in [K-1]$, $s_{q+1}\geq s_{q} $, where $s_{q}$ is defined as
\begin{align}
s_q \vcentcolon= c_{q}-c_{q-1}, 
\label{eq:y in outer bound}
\end{align}
where $c_{q}$ was defined in~\eqref{eq:cq in outer bound}.
\end{lem}
Lemma~\ref{lem:monotony of Y} is key in performing Fourier~Motzkin elimination of $x_{q}$ and $x_{q-1}$ in~\eqref{eq:step 2 final ineq} for each $q\in[K]$, at the end of which we obtain the bound in~\eqref{eq:Rcen outer bound} -- see Appendix~\ref{sec:outer rest of the proof} for details.

The bound in~\eqref{eq:Rcen outer bound} 
is a family of straight lines parameterized by $q\in[K]$.
The lines for $q=t$ and $q=t-1$ intersect at the point in~\eqref{eq:corners in outer bound}
because the coefficients $c_t$ in~\eqref{eq:cq in outer bound} decreases monotonously in $t\in [K-1]$. 
%
This concludes the proof.

\subsection{Remarks}
\label{sub:converse remark}
\paragraph{On the equivalence of our converse bound and other bounds that appeared in the literature after we made available online our works in~\cite{ontheoptimality,ourisitinnerbound}}
\label{rem:we-did-the-converse-first-equivalence}
In~\cite{exactrateuncoded}, the authors propose a genie-aided converse bound to arrive to our very same inequality  in~\eqref{eq:original uncycle}, where~\eqref{eq:original uncycle} was originally derived in~\cite{ontheoptimality,ourisitinnerbound} by leveraging the IC acyclic converse bound. These two approaches are completely equivalent. 
Firstly, the IC acyclic converse bound can be proved by providing genie information to the receivers in the acyclic set. 
Secondly, the following steps in~\cite{exactrateuncoded} are also the same as in our original work~\cite{ontheoptimality,ourisitinnerbound}, namely, 
summing together all the inequalities, 
bounding the load by the new variables $(x_{t}: t\in [0:K])$ defined in~\eqref{eq:defxi}, and 
eliminating the new variables to get the final converse bound. 
The only difference is that 
we use Fourier-Motzkin elimination to eliminate the new variables, while the authors in~\cite{exactrateuncoded} treat the new variables as a probability mass function and optimized the bound over all probability mass functions (see~\eqref{eq:casefilesize}) with a given constraint on the first moment (see~\eqref{eq:case3cachesize}).

\paragraph{Generalization of our converse bound to the case of average load or asymmetric settings}
\label{rem:we-did-the-converse-first-average}
\begin{subequations}
Our proposed converse bound trivially generalizes to different memory sizes or to different file sizes or to average load.
Let the cache size of user $i\in[K]$ be $M_{i} B$ bits, and the length of file $j\in[N]$ be $L_j B$ bits.
We have
\begin{align}
  &R_{\textrm{c,u}}^{\star}
  \geq R_{\textrm{c,low}},
\label{eq:rem:Rclow general}
\end{align}
where $R_{\textrm{c,low}}$ is further lower bounded as
\begin{align}
  &\sum_{\mathbf{d}\in[N]^K} \ \Pr[\mathbf{d}] R_{\mathbf{d}} \leq R_{\textrm{c,low}},
 \ \text{(for average load), or}
\label{eq:rem:Rclow ave}
\\
  &R_{\mathbf{d}} \leq R_{\textrm{c,low}},
 \ \text{(for worst-case load),}
\label{eq:rem:Rclow worst}
\end{align}
where we optimize over the lengths of the subfiles subject to
\begin{align}
  &
\sum_{\mathcal{W}\subseteq [K]} |F_{j,\mathcal{W}}| = L_j B, \ \forall j\in[N], \ \text{(file length)}, 
\label{eq:rem:Rclow file}
\\&
\sum_{j\in[N]} \sum_{\mathcal{W}\subseteq [K] : u\in\mathcal{W}} |F_{j,\mathcal{W}}| \leq M_u B, \ \forall u\in[K], \ \text{(cache size)},
\label{eq:rem:Rclow cache}
\\&
\sum_{i\i\mathcal{N}(d)} \sum_{\Wc_{i}\subseteq[K]\backslash\{u_{1},\ldots,u_{i}\}} \negmedspace\negmedspace\negmedspace\negmedspace |F_{d_{u_{i}},\Wc_{i}}|  \leq R_{\mathbf{d}}, \notag
 \text{ (acyclic IC bound)}, 
\\&
\label{eq:rem:Rclow demand}
\end{align}
\label{eq:rem:Rclow}
where the demand vector $\mathbf{d}\in [N]^K$ in~\eqref{eq:rem:Rclow demand} has $\mathcal{N}(\mathbf{d})$ distinct entries, and where the $\mathbf{u}$ is a permutation of the sub-vector of $\mathbf{d}$ with distinct entries.

The bound in~\eqref{eq:rem:Rclow} is a linear program with 
$\min\{N,K\}! \times N^K+N+K+1 $ constraints in $N2^K+1$ variables, which becomes computationally intense to evaluate when $K$ is large. The symmetry of the caching problem (i.e., invariance to relabeling either the files or the users) enabled us to: 
(i) combine  together the $N$ bounds in~\eqref{eq:rem:Rclow file} in the single constraint like in~\eqref{eq:casefilesize}, 
(ii) combine  together the $K$ bounds in~\eqref{eq:rem:Rclow cache} in the single constraint like in~\eqref{eq:case3cachesize}, and 
(iii) combine together the $\min\{N,K\}! \times N^K$ bounds in~\eqref{eq:rem:Rclow demand} in the single constraint like in~\eqref{eq:case3inequal},
without apparently any loss in optimality.
\end{subequations}


\section{Achievable Bound for Centralized Cache-aided Systems with Uncoded Cache Placement}
\label{sec:centralized inner bound}

In this section, we give an alternate proof to the one in~\cite{exactrateuncoded} that our converse bound in Theorem~\ref{thm:centralized outer bound} is indeed achievable. 

\subsection{Theorem Statement}
\begin{thm}
\label{thm:optimality of cMAN}
The converse bound in Theorem~\ref{thm:centralized outer bound} is achievable by the MAN uncoded cache placement and a delivery scheme based on the IC achievable scheme in Theorem~\ref{thm: novel index coding inner bound}.
\end{thm}
The proof of Theorem~\ref{thm:optimality of cMAN} is given in Section~\ref{sub:achiev}.

\begin{thm}[Novel Achievable Scheme for the General Index Coding Problem]
\label{thm: novel index coding inner bound} 
\begin{subequations}
A non-negative rate tuple $\mathbf{R} \vcentcolon=(R_{1},\ldots,R_{N^{\prime}})$ is achievable
for the IC problem $\big( (\Ac_{j},\Dc_{j}) : j\in[K^{\prime}] \big)$ defined in Section~\ref{sec:model:IC}
if 
\begin{align}
&\mathbf{R}\in
\bigcap_{j\in [K^{\prime}]} \quad
\bigcup_{\small
\Kc_{j}:
\Dc_{j}\subseteq \Kc_{j}\subseteq [N^{\prime}]\backslash \Ac_{j}
}
\mathscr{R}(\Kc_{j}|\Ac_{j},  \Dc_{j}),
\label{eq:novel 1}
\\
&\mathscr{R}(\Kc|\Ac,\Dc)
\vcentcolon= \bigcap_{\Jc : \Jc\subseteq\Kc,  \Dc\cap\Jc\not=\emptyset}
\left\{ \sum_{i\in\Jc}R_{i}< \kappa_{\Jc} \right\},
\label{eq:novel 2}
\end{align}
where in~\eqref{eq:novel 2} $\kappa_{\Jc}$ is defined as
\begin{align}
\kappa_{\Jc}&\vcentcolon=
I\Big(
  (U_{i} : i\in\Jc);  \mathbf{V}
\big|
(U_{i} : i\in\Ac_{j} \cup \Kc_{j}\setminus\Jc )
\Big),
\label{eq:wJ}
\\
\mathbf{V} &\vcentcolon=\big(  V_{\Pc} : \Pc\subseteq[N^{\prime}]\big),
\label{eq:boldX}
\\ V_{\Pc} &\ \text{a function of} \ \big(U_{i}:i\in\Pc\big), \ \forall \Pc\subseteq[N^{\prime}], \label{eq:X function of U}
\end{align}
for some independent auxiliary random variables $(U_{i} : i\in[N^{\prime}])$ and such that
\begin{align}
H\big(
\mathbf{V} 
\big|
\big(U_{i}:i\in\Ac_{j}\big)
\big) <  1,
\ \forall j\in [K^{\prime}]. 
\label{eq:H(X)<c}
\end{align} 
\end{subequations}
\end{thm}
The proof of Theorem~\ref{thm: novel index coding inner bound} is given in Appendix~\ref{sec:Proof for IC}. 
  Note that the cardinality of the auxiliary random variables $U_{i}$, $i\in[N^{\prime}]$, can be bounded as in~\cite[Section 4.2]{hanspaper} (in particular, let $p=1$ and $\mathcal{A}(i)=\mathcal{X}$ in~\cite[Theorem 4.2]{hanspaper}), thus leading to $|\mathcal{U}_{i}|<|\mathcal{X}|+K^{\prime}$.
In addition, since $V_{\Pc}$, $ \Pc\subseteq [N^{\prime}]$, is a function of  $\big(U_{i}:i\in\Pc\big)$, we have $|\mathcal{V}_{\Pc}| \leq \prod_{i\in \Pc} |\mathcal{U}_i|$.

\begin{cor}
\label{cor outperform composite coding}
The composite (index) coding region in Theorem~\ref{thm: thm2 composite coding}
is included in our novel Theorem~\ref{thm: novel index coding inner bound}.
\end{cor}
The proof of Corollary~\ref{cor outperform composite coding} is given in Appendix~\ref{sec:proof of cor 1}.  

\subsection{Proof of Theorem~\ref{thm:optimality of cMAN}}
\label{sub:achiev}
For centralized cache-aided systems under the constraint of uncoded cache placement, 
the claim of Theorem~\ref{thm:optimality of cMAN} is true for $N\geq K$ because the converse bound in~\eqref{eq:corners in outer bound} coincides with the MAN scheme in~\eqref{eq:cMANloadupper}.
For $N < K$, Theorem~\ref{thm: novel index coding inner bound} can be used to achieve~\eqref{eq:corners in outer bound}, as showed next.

We use the same placement phase as MAN for $M=t\frac{N}{K}$, for $t\in[0:K]$, so that the delivery phase is equivalent to an IC problem with $K$ users in which each sub-file $F_{i,\Wc}$, for $i\in \mathcal{N}(\mathbf{d})$, $\Wc\subseteq[K]$ and $|\Wc|=t$, is an independent message, 
and where the desired message and side information sets are given by~\eqref{eq:dk def for caching} and~\eqref{eq:ak def for caching}, respectively.
Note that the message rates in this equivalent IC problem are identical by construction and the number of messages for the worst case-load is $N^{\prime} = \min(N,K) \binom{K}{t}$.
%

In Theorem~\ref{thm: novel index coding inner bound}, following Example~\ref{ex:Insufficiency on Composite Coding},
we let $\Kc_{j}=\Dc_{j}$ for $j\in [K]$, 
we represent $F_{i,\Wc}$ as a binary vector of length $B/\binom{K}{t}$~bits (assumed to be an integer without loss of generality) and we let the corresponding random variable $U$  be equal to the message. 
We also let $V_{\Pc}$ be non zero only for the linear combinations of messages sent by the MAN scheme in~\cite{dvbt2fundamental}.
From~\eqref{eq:novel 2},  for each set $\Jc \subseteq \{F_{d_k,\Wc}:k\notin \Wc\}$, we have $|\Jc|R_\text{sym}<|\Jc| H(U)$.
With this we have $R_\text{sym} = H(U) = B/\binom{K}{t}$ and
\begin{align}
1 = H(X) = B \frac{\binom{K}{t+1}-\binom{K-\min(N,K)}{t+1}}{\binom{K}{t}},
\end{align}
so the symmetric rate is
\begin{align}
R_\text{sym} 
=  \frac{1}{\binom{K}{t+1}
-\binom{K-\min(N,K)|}{t+1}} 
\end{align}
Each receiver in the original caching problem is interested in recovering $\binom{K}{t}$ messages/subfiles, 
or one file of $B$ bits,
thus the `sum-rate rate' delivered to each user is
\begin{align}
R_\text{sum-rate} 
= \frac{\binom{K}{t}}{\binom{K}{t+1}
-\binom{K-\min(N,K)}{t+1}} 
\end{align}
The load in the caching problem is the number of transmissions (channel uses) needed to deliver one file to each user, thus the inverse of $R_\text{sum-rate}$  
 indeed corresponds to the load in~\eqref{eq:cYMAloadupper}.

\subsection{Remarks}
\label{sub:achiev remark}
\paragraph{On the interpretation on the MAN scheme as source coding with side information} 
Our proof of Theorem~\ref{thm:optimality of cMAN} uses Theorem~\ref{thm: novel index coding inner bound} and gives an interpretation of the achievable scheme proposed in~\cite{exactrateuncoded} via the framework of \emph{source coding with side information}. 
Our novel IC approach has the advantage that it applies to any IC problem, and is not limited to binary linear codes for the specific MAN placement as that of~\cite{exactrateuncoded} when applied to the caching problem.

\paragraph{On general achievable regions for IC}
The composite (index) coding region in Theorem~\ref{thm: thm2 composite coding} is included in our novel Theorem~\ref{thm: novel index coding inner bound}, as shown in Appendix~\ref{sec:proof of cor 1}. The proof of Corollary~\ref{cor outperform composite coding} can be found in Appendix~\ref{sec:proof of cor 1}. 
In addition, from the proof of Corollary~\ref{cor outperform composite coding}, we see that the rate region achieved by composite (index) coding can be realized by linear coding.

The work in~\cite{liu2017ondistributed} improves on the composite (index) coding scheme in Theorem~\ref{thm: thm2 composite coding} by further rate-splitting. In Remark~\ref{rem:rate splitting} in Appendix~\ref{sec:Proof for IC} we discuss how the same idea can be used to improve on Theorem~\ref{thm: novel index coding inner bound}.
Finally, in Example~\ref{ex:Insufficiency on Composite Coding} in Appendix~\ref{sec:Proof for IC} we give an example to show that our Theorem~\ref{thm: novel index coding inner bound} strictly improves on~\cite{liu2017ondistributed}. 
It thus appears that our region in Theorem~\ref{thm: novel index coding inner bound} is the largest known achievable region to date of the general IC based on random coding.

\paragraph{On the extension to decentralized systems}
By directly extending our proposed achievable scheme in Theorem~\ref{thm:optimality of cMAN} and converse bound in Theorem~\ref{thm:centralized outer bound} to decentralized systems, we can prove that under the constraint that each user randomly, uniformly and independently chooses $MB$ bits of the $N$ files to store in its local cache, the optimal load can be achieved by the novel IC achievable bound in Theorem~\ref{thm: novel index coding inner bound}; this result was shown in~\cite{ontheoptimality}.
More precisely, for the converse part, since each user randomly, uniformly and independently chooses $MB$ bits of the $N$ files to store, by a Low-of-Large-Numbers type reasoning, the length of each subfile does not deviate much from its mean when the file size is large. Hence, we can use the technique proposed in Section~\ref{sub:converse proof} to find the largest acyclic sets of subfiles for each possible demand vector. For the achievability part, we can use the proposed delivery scheme in Section~\ref{sub:achiev} for $K$ rounds, where in each round we transmit the subfiles known by exactly $t\in [0:K]$ users. 
The details can be found in the first author's Ph.D. thesis~\cite{kai-defense}.

\section{Conclusion} 
\label{sec:conclusions}

In this paper we investigated the coded caching problem with uncoded cache placement by leveraging its connection to the index coding problem.
We first derived a converse bound on the worst-case load of cache-aided systems under the constraint of uncoded cache placement by cleverly combining many acyclic index coding converse bounds derived by considering different demands in the caching problem. 
When there are more users than files, we proved that the load of the MAN scheme coincides with the proposed converse bound. In the remaining cases, our converse bound is attained by using the MAN placement phase and a delivery phase based on novel index coding scheme. 
The proposed novel index coding achievable scheme is based on distributed source coding and is shown to strictly improves on the well known composite (index) coding achievable bound and is, to the best our knowledge, the best random coding achievable bound for the general IC problem to date. 

The present work parallels the recent results in~\cite{exactrateuncoded}. Our main contribution compared to~\cite{exactrateuncoded} is to build on the connection among the caching problem with uncoded cache placement and the index coding problem.


\appendices

\section{Proof of Lemma~\ref{lem:For-each-graph}}
\label{sec:proof of lemma1}

For a $\mathbf{u}=(u_{1},u_{2},\ldots,u_{\min(K,N)})$, we say that sub-files/nodes $F_{d_{u_{i}},\Wc_{i}}$, for all $\Wc_{i}\subseteq[K]\backslash \{u_{1},\dots,u_{i}\}$, are in level $i$. It is easy to see each node in level $i$ only knows the sub-files $F_{j,\Wc}$ where $u_{i}\in\Wc$. So each node in level $i$ knows neither the sub-files in the same level, nor the sub-files in the higher levels. As a result, in the proposed set there is no sub-set containing a directed cycle.

\section{Proof of Lemma~\ref{lem:monotony of Y}}
\label{sec:proof of lemma2}

Recall that 
\begin{align}
\Yq &=\displaystyle c_q-c_{q-1},
\label{eq:def of y again}\\
\textrm{and }c_q&=\frac{\binom{K}{q+1}-\binom{K-\min(K,N)}{q+1}}{\binom{K}{q}}\nonumber\\
&=\frac{\binom{K-1}{q}+\cdots+\binom{K-\min(K,N)}{q}}{\binom{K}{q}}\label{eq:derivation of cq}.
\end{align}
Focus on the first term of $\Yq$ in~\eqref{eq:def of y again}, we have
\begin{align}
 & c_q=\frac{\binom{K-1}{q}+\cdots+\binom{K-\min(K,N)}{q}}{\binom{K}{q}}= \frac{K-q}{K}+\cdots+\nonumber\\
 &\frac{(K-q)\times...\times(K-q-\min(K,N)+1)}{K\times...\times(K-\min(K,N)+1)}.
\label{eq:first term of y}
\end{align}
For the second term of $\Yq$
\begin{align}
 & c_{q-1}=\frac{\binom{K-1}{q-1}+\cdots+\binom{K-\min(K,N)}{q-1}}{\binom{K}{q-1}} =\frac{K-q+1}{K}+\cdots+\nonumber\\
 &\frac{(K-q+1)\times\cdots \times(K-q-\min(K,N)+2)}{K\times\cdots\times(K-\min(K,N)+1)}.\label{eq:second term of y}
\end{align}
By taking~\eqref{eq:first term of y} and~\eqref{eq:second term of y} into~\eqref{eq:def of y again}, we  finally obtain
\begin{align}
&\Yq =  -\frac{1}{K}-\frac{2(K-q-1)}{K(K-1)}
-\cdots
\\&
-\frac{\min(K,N)(K-q)\times\cdots\times(K-q-\min(K,N)+2)}{K\times\cdots\times(K-\min(K,N)+1)}.\label{eq:sum form of y}
\end{align}
Since each negative term in~\eqref{eq:sum form of y} is monotone increasing with $q$,
 it is easy 
 to check that for any $q\in[K-1]$, $s_{q+1}\geq \Yq $.

\section{Proof of~\eqref{eq:Rcen outer bound}}
\label{sec:outer rest of the proof}

From~\eqref{eq:casefilesize} (i.e., the fact that $(x_0, \ldots, x_K)$ as defined in~\eqref{eq:defxi} can be interpreted as a probability mass function), we have
\begin{align}
& (c_{q}-q\Yq)(x_{q}+x_{q-1})
 = (c_{q}-q\Yq)\left(1- \sum_{i\in[0:K] \backslash\{q-1,q\}} x_{i}\right)
 \label{eq:from file size},
\end{align}
where $\Yq $ is given in~\eqref{eq:def of y again}. 
By the Lemma~\ref{lem:monotony of Y}, for any $q\in [K-1]$, $s_{q+1}\geq \Yq$. Since $s_{K}\leq 0$, we have $\Yq\leq 0$ for all $q\in [K]$.
From~\eqref{eq:case3cachesize}, we have
\begin{align}
& \Yq \big((q-1)x_{q-1}+qx_{q}\big)\geq 
  \Yq \left( \frac{KM}{N}-\sum_{i\in[0:K] \backslash\{q-1,q\}}  ix_{i}\right).
\label{eq:from cache size}
\end{align}
By summing~\eqref{eq:from file size} and~\eqref{eq:from cache size} we get
\begin{align}
  & c_{q-1}x_{q-1}+c_{q}x_{q}
 \geq \nonumber \\ &\Yq \ \frac{KM}{N} 
 +c_{q}
 -\Yq \ q
  + \sum_{i\in[0:K]:i\neq q-1,q}  (-c_{q}+(q-i)\Yq)x_{i}.
\label{eq:from the sum}
\end{align}
Next, we substitute~\eqref{eq:from the sum} into~\eqref{eq:step 2 final ineq} and get
\begin{align}
&R_{\textrm{c,u}}^{\star} 
\geq  \frac{\Yq KM}{N}
+c_{q}
-\Yq \ q
+\sum_{i\in[0:K]} w_{q,i}x_{i},
\label{eq:ineq for each q}
\\
&w_{q,i} \vcentcolon=  c_{i}-c_{q}
  +(q-i)\Yq .
 \label{eq:W(K,N,q)}
\end{align}
Note that when $i\in \{q,q-1\}$ we have $w_{q,i}=0$.
It remains to prove that for each $i\in[0:K]$ we have $w_{q,i}\geq 0$.
For any $q\in [K]$ and $i\in[0:K-1]$ we have
\begin{align}
w_{q,i+1}- w_{q,i}= & \frac{s_{i+1}}{N}-\frac{\Yq }{N}.\label{eq:monotony of W}
\end{align}
From Lemma~\ref{lem:monotony of Y} and~\eqref{eq:monotony of W}, it can be seen that for any $q\in [K]$ and $i\in[0:K-1]$, if $i\leq q-1$, $w_{q,i+1}\leq w_{q,i}$ and if $i\geq q-1$, $w_{q,i+1}\geq w_{q,i}.$ Furthermore, $w_{q,i}=0$ for $i\in \{q,q-1\}$. Hence, for each $i\in[0:K]$, $w_{q,i}\geq 0$. 
As a result we have
\begin{align}
&R_{\textrm{c,u}}^{\star}
\geq  \frac{\binom{K}{q+1}-\binom{K-\min(K,N)}{q+1}}{\binom{K}{q}}
+\Yq \left( \frac{KM}{N}-q\right),
\end{align}
which proves bound given in~\eqref{eq:Rcen outer bound}.

\section{Proof of Theorem~\ref{thm: novel index coding inner bound}}
\label{sec:Proof for IC}

We introduce here a novel IC achievable scheme based on coding for the Multi-Access Channel (MAC) with correlated messages~\cite{hanspaper}, Slepian-Wolf coding~\cite{slepianwolf}, and non-unique decoding~\cite{nonunique}. 
At a very high level, the proposed scheme can be described as follows, where the terminology and notation are as in Section~\ref{sec:model:IC in}. In the encoding stage, we generate a sequence  for each message and then generate a composite function for each subset of sequences. In the decoding stage, we choose a set $\Kc_j$ such that $\Dc_j \subseteq  \Kc_j$ for each user $j$. From all the received composite functions and the side information of user $j$,
we let user $j$ uniquely decode the messages in $\Dc_j$, non-uniquely decode the messages in $\Kc_j\setminus \Dc_j$, and treat the other messages as interference. 
We then prove that the rate region of the proposed scheme not only strictly includes the region achieved by composite (index) coding in Theorem~\ref{thm: thm2 composite coding} but it also strictly outperforms the improved version of Theorem~\ref{thm: thm2 composite coding} from~\cite{liu2017ondistributed}. 
Our scheme differs from Theorem~\ref{thm: thm2 composite coding} in the following aspects:
\begin{enumerate}
\item 
In the composite (index) coding scheme, decoder $j \in[K^{\prime}]$  recovers uniquely the messages in $\Kc_{j}$, while in our proposed scheme decoder $j \in[K^{\prime}]$ uniquely recovers only the desired messages indexed by $\Dc_{j}$ and non-uniquely the non-desired indexed by $\Kc_{j}\backslash \Dc_{j}$. So in~\eqref{eq:composite 2} the intersection is taken over all of $\Jc \subseteq \Kc$ while in~\eqref{eq:novel 2} the intersection is taken over all of $\Jc\subseteq \Kc$ such that $\Dc \cap \Jc=\emptyset$. 

\item  
In the composite (index) coding scheme decoder $j \in[K^{\prime}]$ treats all the messages in $[N^{\prime}]\setminus \Kc_j$ as noise and the correlation among composite indices is not considered. Thus decoder $j$ only uses  the composite indices $(W_{\Jc} : \Jc \subseteq \Kc_{j}\cup \Ac_{j})$ to decode all the messages in $\Kc_{j}$. Instead, in our proposed scheme, decoder $j$ treats all the messages in $[N^{\prime}]\setminus \Kc_j$ as interference. By leveraging the correlation among all the composite functions, we let decoder $j$ cancel the interference of $[N^{\prime}]\setminus \Kc_j$. 

For instance, in Example~\ref{ex:Insufficiency on Composite Coding} at the end of this section, user~$3$ knows messages indexed by $\{5,6\}$  and demands message~$3$, i.e., $\Ac_3=\{5,6\}$  and $\Dc_3=\{3\}$. In the proposed scheme, which is proven to be optimal for this example, user~$3$ uses all the transmitting composite indices to recover message~$3$ and cancel the interference of the messages indexed by $\{1,2,4\}$ without decoding those messages. However, if we use composite (index) coding, user~$3$ can only use the composite indices $W_{\Jc}$ if and only if $\Jc$ is a subset of $\Kc_3\cup \Ac_3$, (e.g., if we set $\Kc_3=\Dc_3=\{3\}$, user $3$ treats all the composite indices $V_{1,3,4}$, $V_{2,4,5}$ and $V_{1,2,6}$  as noise; else if we set $\Kc_3\supset\Dc_3$, user $3$ should exactly recover all messages in $\Kc_3$ which includes some messages not demanded by user $3$; in both cases, the composite coding can not achieve the converse bound). This is the main reason why composite (index) coding is not optimal in this example and why our proposed scheme outperforms  composite coding.
 
\end{enumerate}

\begin{IEEEproof}
To clarify the notations, we use different symbols for 
transmitted messages or known messages (nothing above), 
uniquely decoded ones (hat above), and 
non-uniquely decoded ones (check above).
\paragraph{Codebook Generation}
Fix a probability mass function
\begin{align}
p_{U_{1},\dots,U_{N^{\prime}}}(u_{1},\dots,u_{N^{\prime}})=p_{U_{1}}(u_{1})\times\cdots\times p_{U_{N^{\prime}}}(u_{N^{\prime}}),
\end{align}
where each random variable $U_{i}$ is defined on the finite alphabet $\mathcal{U}_i$ for $i\in[N^{\prime}]$,
and functions
\begin{align}
f_{\mathcal{P}}
: \prod_{i\in \mathcal{P}} \mathcal{U}_i \to  \mathcal{V}_{\mathcal{P}},
\quad \forall \mathcal{P}\subseteq[N^{\prime}],
\end{align}
for some finite alphabets $\mathcal{V}_{\mathcal{P}}$ for $ \mathcal{P}\subseteq[N^{\prime}]$.

For each $i\in[N^{\prime}]$, randomly and independently generate $|\mathcal{X}|^{nR_{i}}$ sequences $u_{i}^{n}(m_{i})$
indexed by $m_{i}\in[|\mathcal{X}|^{nR_{i}}]$, each according to $\prod_{t=1}^{n}p_{U_{i}}(u_{i,t})$.
For each $\mathcal{P}\subseteq[N^{\prime}]$, let 
$v_{\mathcal{P}}^{n} \vcentcolon= (v_{\mathcal{P},1},\dots,v_{\mathcal{P},n})$
and $v_{\mathcal{P},t}  =f_{\mathcal{P}}\big((u_{i,t}:i\in\mathcal{P})\big) \in \mathcal{V}_{\mathcal{P}}$ where $t\in[n]$. 

Randomly and independently assign an index $g\in[|\mathcal{X}|^{n}]$ to
each collection of sequences $(v_{\mathcal{P}}^{n}:\mathcal{P}\subseteq[N^{\prime}])$ according to a uniform
probability mass function over $[|\mathcal{X}|^{n}]$. The sequences with the same index $g$ are said to form
bin $\mathcal{B}(g)$. We also indicate $g = \mathsf{bin}(v_{\mathcal{P}}^{n}:\mathcal{P}\subseteq[N^{\prime}])$, the index of the bin of $v_{\mathcal{P}}^{n}$. 

The codebook so generated is revealed to all the decoders.

\paragraph{Encoding} 
Given messages $(m_1,\ldots,m_{N^{\prime}})$, the encoder produces 
$(u_{1}^{n}(m_{1}), \ldots, u_{N^{\prime}}^{n}(m_{N^{\prime}}))$ based on which it computes
$(v_{\mathcal{P}}^{n}:\mathcal{P}\subseteq[N^{\prime}])$ and
eventually transmits $g = \mathsf{bin}(v_{\mathcal{P}}^{n}:\mathcal{P}\subseteq[N^{\prime}])$ to all the decoders.

\paragraph{Decoding}
Fix $\mathcal{K}_{j}$ where $\Dc_{j}\subseteq\mathcal{K}_{j}$
and $\mathcal{K}_{j}\cap\Ac_{j}=\emptyset$ for each receiver $j\in [K^{\prime}]$. 
Decoding proceeds in two steps.

Decoding~Step~1:
Since receiver $j\in [K^{\prime}]$ has messages $(m_{i}:i\in \Ac_{j})$ as side information, it also knows 
$(u_{i}^{n} : i\in\Ac_{j})$ and $(v_{\mathcal{P}}^{n} : \mathcal{P}\subseteq\Ac_{j})$.
Upon receiving $g = \mathsf{bin}(v_{\mathcal{P}}^{n}:\mathcal{P}\subseteq[N^{\prime}])$, receiver $j\in [K^{\prime}]$ estimates the sequences 
$(\widehat{v}_{\mathcal{P}}^{n}:\mathcal{P}\subseteq[N^{\prime}], \ \mathcal{P}\nsubseteq\Ac_{j})$
as the unique sequences satisfying
\begin{align}
\big(
(\widehat{v}_{\mathcal{P}}^{n}:\mathcal{P}\subseteq[N^{\prime}], \ \mathcal{P}\nsubseteq\Ac_{j}),
(v_{\mathcal{P}}^{n} : \mathcal{P}\subseteq\Ac_{j})
\big)
\in \mathcal{B}(g);
\end{align} 
if none or more than one are found, it picks one uniformly at random within $\mathcal{B}(g)$.

Decoding~Step~2:
Receiver $j\in [K^{\prime}]$ then
uses the found $(\widehat{v}_{\mathcal{P}}^{n}:\mathcal{P}\subseteq[N^{\prime}], \ \mathcal{P}\nsubseteq\Ac_{j})$,
and the side information to decode all messages in $\mathcal{K}_{j}$, but only those in $\Dc_{j}$ uniquely, that is, it 
finds 
a unique tuple $(\widehat{m}_{i}:i\in\mathcal{D}_{j})$ 
and some tuple $(\check{m}_{i}:i\in\mathcal{K}_{j}\backslash\mathcal{D}_{j})$ such that
\begin{subequations}
\begin{align}
\Big(
&
\big(u_{i}^{n}(m_{i}):i\in \Ac_{j}\big), \ 
\big(u_{i}^{n}(\widehat{m}_{i}):i\in\mathcal{D}_{j}\big), \ 
\\&
\big(u_{i}^{n}(\check{m}_{i}):i\in\mathcal{K}_{j}\backslash\mathcal{D}_{j}\big), \
\big(\widehat{v}_{\mathcal{P}}^{n}:\mathcal{P}\subseteq[N^{\prime}], \ \mathcal{P}\nsubseteq\Ac_{j}\big), \
\\&
\big(v_{\mathcal{P}}^{n} : \mathcal{P}\subseteq\Ac_{j}\big)
\Big)
\\&
\in T_{\varepsilon}^{(n)}\Big(
\big(U_{i}:i\in \Ac_{j}\cup \mathcal{K}_{j}\big),
\big(V_{\mathcal{P}}:\mathcal{P}\subseteq[N^{\prime}]\big)
\Big);
\end{align}
if none or more than one $(\widehat{m}_{i}:i\in\mathcal{D}_{j})$ are found, it picks one uniformly at random.
\end{subequations}


\paragraph{Error Analysis}
For each decoder $j\in [K^{\prime}]$ and $\mathcal{J}\subseteq\mathcal{K}_{j}$ where $\mathcal{J}\cap\Dc_{j}\neq\emptyset$, we define the error events in~\eqref{eq:errorAll} at the top of the next page. 

\begin{figure*}
\begin{subequations}
\begin{align}
\mathcal{E}_{1} &\vcentcolon= \big\{\big((U_{i}^{n}(M_{i}):i\in[N^{\prime}]),(V_{\mathcal{P}}^{n}:\mathcal{P}\subseteq[N^{\prime}])\big) \notin T_{\varepsilon}^{(n)}\big((U_{i}:i\in [N^{\prime}]),({V}_{\mathcal{P}}^{n}:\mathcal{P}\subseteq[N^{\prime}])\big)\big\},
\label{eq:error1}
\\ 
\mathcal{E}_{2,j} &\vcentcolon=  \big\{\textrm{there exists }(\widehat{v}_{\mathcal{P}}^{n}:\mathcal{P}\subseteq[N^{\prime}]\textrm{ and }\mathcal{P}\nsubseteq\Ac_{j})\in T_{\varepsilon}^{(n)}\big((V_{\mathcal{P}}:\mathcal{P}\subseteq[N^{\prime}]\textrm{ and }\mathcal{P}\nsubseteq\Ac_{j})|(U_{i}^{n}:i\in \Ac_{j})\big) 
\nonumber\\
&\textrm{ such that }\big((\widehat{v}_{\mathcal{P}}^{n}:\mathcal{P}\subseteq[N^{\prime}]\textrm{ and }\mathcal{P}\nsubseteq\Ac_{j}),(V_{\mathcal{P}}^{n}:\mathcal{P}\subseteq\Ac_{j})\big)\in\mathcal{B}(G)\textrm{ and }(\widehat{v}_{\mathcal{P}}^{n}:\mathcal{P}\subseteq[N^{\prime}]
 \nonumber\\
&\textrm{ and }\mathcal{P}\nsubseteq\Ac_{j})\neq(V_{\mathcal{P}}^{n}:\mathcal{P}\subseteq[N^{\prime}]\textrm{ and }\mathcal{P}\nsubseteq\Ac_{j})\big\}, \textrm{ where }G\textrm{ is the random index of } g,
\label{eq:error2}
\\ 
\mathcal{E}_{j,\mathcal{J}} &\vcentcolon= \big\{\textrm{there exists }\widehat{m}_{i}\neq M_{i}\textrm{
where }i\in\mathcal{J} \textrm{ such that }\big((U_{i}^{n}(M_{i}):i\in\mathcal{K}_{j}\cup\Ac_{j}\setminus\mathcal{J}),(U_{i}^{n}(\widehat{m}_{i}):i\in\mathcal{J}),  
\nonumber\\
&(V_{\mathcal{P}}^{n}:\mathcal{P}\subseteq[N^{\prime}])\big)\in T_{\varepsilon}^{(n)}\big((U_{i}:i\in \mathcal{K}_{j}\cup\Ac_{j}),(V_{\mathcal{P}}:\mathcal{P}\subseteq[N^{\prime}])\big)\big\}.
\label{eq:error3}
\end{align}
\label{eq:errorAll}
\end{subequations}
\end{figure*}

For decoder $j$, the probability of error at decoder $j$ denoted by $\Pr(\mathcal{E}(j))$ can be upper bounded by
\begin{subequations}
\begin{align}
\Pr\big(\mathcal{E}(j)\big)
  &\leq\Pr(\mathcal{E}_{1})
\label{eq:error userj 1}
\\&+\Pr\big(\mathcal{E}_{1}^{c}\cap\mathcal{E}_{2,j}|\mathcal{B}(1)\big)
\label{eq:error userj 2}
\\&+\sum_{\mathcal{J}\subseteq\mathcal{K}_{j}:\mathcal{J}\cap\Dc_{j}\neq\emptyset}\Pr\left(\mathcal{E}_{j,\mathcal{J}}\cap\mathcal{E}_{1}^{c}\cap\mathcal{E}_{1,j}^{c}\right).
\label{eq:error userj 3}
\end{align}
\label{eq:error userj all}
\end{subequations}
We now bound each term in~\eqref{eq:error userj all}. 
By LLN, the term in~\eqref{eq:error userj 1} vanishes 
as $n\rightarrow\infty$. Next, for the term in~\eqref{eq:error userj 2} we have
\begin{align}
&\Pr\big(\mathcal{E}_{1}^{c}\cap\mathcal{E}_{2,j}|\mathcal{B}(1)\big)\leq\sum_{(u_{i}^{n}:i\in [N^{\prime}])}\Pr\Big\{U_{i}^{n}=u_{i}^{n},i\in[N^{\prime}]\nonumber\\
&\Big|\Big(f_{\mathcal{P}}\big((U_{i}^{n}:i\in\mathcal{P})\big):\mathcal{P}\subseteq[N^{\prime}]\Big)\in\mathcal{B}(1)\Big\}q_{(u_{i}^{n}:i\in[N^{\prime}])}
\label{eq:E_2,j}
\end{align}
where 
\begin{align}
&q_{(u_{i}^{n}:i\in[N^{\prime}])}\vcentcolon=\nonumber\\
&\Pr\Big\{\big((\widehat{v}_{\mathcal{P}}^{n}:\mathcal{P}\subseteq[N^{\prime}]\textrm{ and }\mathcal{P}\nsubseteq\Ac_{j}), (x_{\mathcal{P}}^n:\mathcal{P} \subseteq \Ac_j)\big)\nonumber\\
&\in\mathcal{B}(1)\textrm{ for some }
(\widehat{v}_{\mathcal{P}}^{n}:\mathcal{P}\subseteq[N^{\prime}]\textrm{ and }\mathcal{P}\nsubseteq\Ac_{j})\in\nonumber\\ &\mathcal{G}_{(u_{i}^{n}:i\in[N^{\prime}])} 
\Big| \Big(f_{\mathcal{P}}\big((U_{i}^{n}:i\in\mathcal{P})\big):\mathcal{P}\subseteq[N^{\prime}]\Big)\in\mathcal{B}(1),\nonumber\\
&U_{i}^{n}=u_{i}^{n} \textrm{ where }i\in[N^{\prime}]\Big\};\nonumber\\
&\mathcal{G}_{(u_{i}^{n}:i\in[N^{\prime}])}\vcentcolon=\Big\{(\widehat{v}_{\mathcal{P}}^{n}:\mathcal{P}\subseteq[N^{\prime}]\textrm{ and }\mathcal{P}\nsubseteq\Ac_{j})\neq \nonumber\\
&\Big(f_{\mathcal{P}}\big((u_i^{n}:i\in \mathcal{P})\big):\mathcal{P}\subseteq[N^{\prime}]\textrm{ and }\mathcal{P}\nsubseteq\Ac_{j}\Big):\nonumber\\
&(\widehat{v}_{\mathcal{P}}^{n}:\mathcal{P}\subseteq[N^{\prime}]\textrm{ and }\mathcal{P}\nsubseteq\Ac_{j})\in \nonumber\\
&T_{\varepsilon}^{(n)}\big((V_{\mathcal{P}}:\mathcal{P}\subseteq[N^{\prime}]\textrm{ and }\mathcal{P}\nsubseteq\Ac_{j})|(u_{i}^{n}:i\in\Ac_{j})\big)\Big\}.
\end{align}
We then focus on $q_{(u_{i}^{n}:i\in[N^{\prime}])}$ to obtain
\begin{align}
&q_{(u_{i}^{n}:i\in[N^{\prime}])}\leq                        \sum_{\stackrel{(\widehat{v}_{\mathcal{P}}^{n}:\mathcal{P}\subseteq[N^{\prime}]\textrm{ and }\mathcal{P}\nsubseteq\Ac_{j})\in}{T_{\varepsilon}^{(n)}\big((X_{\mathcal{P}}:\mathcal{P}\subseteq[N^{\prime}]\textrm{ and }\mathcal{P}\nsubseteq\Ac_{j})|(u_{i}^{n}:i\in\Ac_{j})\big)}}                   \nonumber\\&     \Pr\Big\{\big((\widehat{v}_{\mathcal{P}}^{n}:\mathcal{P}\subseteq[N^{\prime}],\mathcal{P}\nsubseteq\Ac_{j}),(x_{\mathcal{P}}^n:\mathcal{P}\subseteq\Ac_{j})\Big)\in\mathcal{B}(1) \Big| \nonumber\\
&\Big(f_{\mathcal{P}}\big((U_{i}^{n}:i\in\mathcal{P})\big):\mathcal{P}\subseteq[N^{\prime}]\Big)\in\mathcal{B}(1),U_{i}^{n}=u_{i}^{n} \forall i\in[N^{\prime}]  \Big\}
\nonumber\\
& \leq |\mathcal{X}|^{n{\displaystyle \left[H\big((V_{\mathcal{P}}:\mathcal{P}\subseteq[N^{\prime}], \mathcal{P}\nsubseteq\Ac_{j})|(U_{i}:i\in\Ac_{j})\big)\right]}}|\mathcal{X}|^{-n}.\label{eq:q(xpn)}
\end{align}
From~\eqref{eq:E_2,j} and~\eqref{eq:q(xpn)} we can see that the term in~\eqref{eq:error userj 1} 
vanishes provided that 
\begin{align}
H\big( (V_{\mathcal{P}}:\mathcal{P}\subseteq[N^{\prime}]\textrm{ and }\mathcal{P}\nsubseteq\Ac_{j})|(U_{i}:i\in \Ac_{j})\big) < 1.
\label{eq:1st with strictly less}
\end{align}
Finally, for 
the term in~\eqref{eq:error userj 3} also vanishes, 
provided that (by the packing lemma~\cite[Lemma~3.1]{elgamalkim})
\begin{align}
\sum_{i\in\mathcal{J}}R_{i} 
 & < I\big((U_{i}:i\in \mathcal{J});(V_{\mathcal{P}}:\mathcal{P}\subseteq[N^{\prime}]
 \label{eq:2nd with strictly less}
 \textrm{ and }
 \\
 &\mathcal{P}\nsubseteq\mathcal{K}_{j}\cup\Ac_{j}\setminus \mathcal{J}),(U_{i}:i\in \mathcal{K}_{j}\cup\Ac_{j}\setminus\mathcal{J})\big)\\
 &= I\big((U_{i}:\in \mathcal{J});(V_{\mathcal{P}}:\mathcal{P}\subseteq[N^{\prime}]\textrm{ and }\nonumber\\
 &\mathcal{P}\nsubseteq\mathcal{K}_{j}\cup\Ac_{j}\setminus\mathcal{J})|(U_{i}:i\in \mathcal{K}_{j}\cup\Ac_{j}\setminus\mathcal{J})\big).
\end{align}
  The inequalities in~\eqref{eq:1st with strictly less} and in~\eqref{eq:2nd with strictly less} are strict; however, by the same argument in the composite-coding achievable region in~\cite[Proposition 6.11]{onthecapacityindex} they can be relaxed to be nonstrict.
\end{IEEEproof}

\begin{rem}
\label{rem:rate splitting}
The work in~\cite{liu2017ondistributed} improves on the composite (index) coding scheme in Theorem~\ref{thm: thm2 composite coding} by splitting each message into non-overlapping and independent sub-message 
\begin{align*}
M_i=\big(M_i\big((\Kc_1,\ldots,\Kc_{K^{\prime}})\big):\Kc_k\subseteq [N^{\prime}]\setminus \Ac_k, \textrm{ for }k\in[K^{\prime}]\big).
\end{align*}
Composite (index) coding is then used to transmit each group of sub-messages, $\{M_i\big((\Kc_1,\ldots,\Kc_{K^{\prime}})\big):i\in [N^{\prime}]\}$.  A drawback of this scheme is that the number of auxiliary variable increases   exponentially with the number of users.

We could also use this message-splitting idea to improve our achievable region in Theorem~\ref{thm: novel index coding inner bound}. If we did so, then the improved version of Theorem~\ref{thm: novel index coding inner bound} would include the improved version of Theorem~\ref{thm: thm2 composite coding}. This message-splitting improvement is quite straightforward and not pursued here. We note that the improvement in performance comes at the expense of a much heaver notation, and--more importantly--a much larger computation burden to evaluate a regions that is already combinatorial in nature.

We would like to stress that our main objective here is to propose a general IC achievable scheme that can achieve the converse bound for the caching problem under the constraint of uncoded cache placement, and at the same time improves on the composite (index) coding. To the best of our knowledge, such general scheme does not exist in the literature. Corollary~\ref{cor outperform composite coding}  and of Theorem~\ref{thm: novel index coding inner bound} show that the current achievable bound in Theorem~\ref{thm: novel index coding inner bound} is sufficient to achieve our objective. In addition, in some special cases such as in the following Example~\ref{ex:Insufficiency on Composite Coding}, we show that even if we do not use rate splitting, our proposed achievable bound is strictly better than the the message-split region in~\cite[SectionIII-B]{liu2017ondistributed}. 
\end{rem}

\begin{example}
\label{ex:Insufficiency on Composite Coding}
\rm

Consider a multiple unicast IC problem with $K^\prime=6$ messages and with
\begin{align*}
& \Dc_{1}=\{1\}, \quad \Ac_{1}=\{3,4\},\\
& \Dc_{2}=\{2\}, \quad \Ac_{2}=\{4,5\},\\
& \Dc_{3}=\{3\}, \quad \Ac_{3}=\{5,6\},\\
& \Dc_{4}=\{4\}, \quad \Ac_{4}=\{2,3,6\},\\
& \Dc_{5}=\{5\}, \quad \Ac_{5}=\{1,4,6\},\\
& \Dc_{6}=\{6\}, \quad \Ac_{6}=\{1,2\}.
\end{align*}

\paragraph*{Composite (Index) Coding Achievable Bound}
In~\cite[Example~1]{liu2017ondistributed} the authors showed that the largest symmetric rate with the composite (index) coding achievable bound in Theorem~\ref{thm: thm2 composite coding} for this problem is $R_\text{sym,cc}=0.2963$. It the same paper, the authors proposed an extension of the composite (index) coding idea (see~\cite[Section III.B]{liu2017ondistributed}) and showed that this extended scheme for this problem gives $R_\text{sym,cc,enhanced}=0.2987$.

\paragraph*{Converse}
Give message $M_5$ as additional side information to receiver~1 so that the new side information set satisfied $\{3,4,5\} \subseteq \Ac_{1}$.
With this side information, in addition to message~1, receiver~1 can decode message~2 and then message~6 for a total of $3$ messages.
Thus 
\begin{align}
3R_\text{sym} \leq \lim_{n\to\infty}\frac{1}{n}H(X^n) \leq  1,
\label{eq:outer example} 
\end{align}
where $R_\text{sym}$ denotes the symmetric rate.
Next we show that $R_\text{sym} \leq 1/3$ is tight. 
{\it This shows the strict sub-optimality of composite (index) coding and its message-split extension.}

\paragraph*{Achievability}
 It is not difficult to see that all users can be satisfied by the transmission of the three coded messages
$X=(M_1\oplus M_3\oplus M_4, \ M_2\oplus M_4\oplus M_5, \  M_1\oplus M_2\oplus M_6)$.

We now map this scheme into a choice of auxiliary random variables in our novel IC scheme in Theorem~\ref{thm: novel index coding inner bound}.
Let
$\Kc_{j}=\Dc_{j}$ for $j\in [6]$, and 
\begin{align*}
  &U_{1} = M_1, \ U_{2} = M_2, \cdots, U_{6} = M_6,
\\&\text{for all $\Pc\subseteq[6]$ set $V_{\Pc}=0$ except for the following}
\\&V_{\{1,3,4\}}=U_{1}\oplus U_{3}\oplus U_{4},
\\&V_{\{2,4,5\}}=U_{2}\oplus U_{4}\oplus U_{5},
\\&V_{\{1,2,6\}}=U_{1}\oplus U_{2}\oplus U_{6},
\end{align*}
Hence,
$
\mathbf{V}=(V_{\{1,3,4\}},V_{\{2,4,5\}},V_{\{1,2,6\}}).
$
From~\eqref{eq:novel 2}, we have that for example the rate bound corresponding to receiver~5 is
\begin{align*}
&R_\text{sym} 
<
I(
U_{5}
; \mathbf{V}
|
U_{1}, U_{4}, U_{6}
)
\\
&= 
I(
U_{5}
;
U_{3},
U_{2}\oplus U_{5},
U_{2}
)
= 
I(
U_{5}
;
U_{2},
U_{3},
U_{5}
)
= 
I(
U_{5}
;
U_{5}
)
\\
&=
H (U_{5})=1/3, 
\end{align*}
and similarly for all the other users.
 As a result, any $R_\text{sym}< 1/3$ is achievable by the proposed scheme based on random coding argument;
by repeating the same argument with random linear codes, $R_\text{sym} \leq 1/3$ is achievable and coincides with the converse bound.
\hfill$\square$
\end{example}

\section{Proof of Corollary~\ref{cor outperform composite coding}}
\label{sec:proof of cor 1}

In general, for a set $\Bc\subseteq[N^{\prime}]$ and for the auxiliary random variables as defined in Theorem~\ref{thm: novel index coding inner bound}, we have
\begin{align}
&
H\big((V_{\Pc}:\Pc\subseteq[N^{\prime}] ) 
\big|
(U_{i} : i\in \Bc) 
\big)
\notag\\&\leq H\big( (V_{\Pc}:\Pc\subseteq[N^{\prime}], \Pc\not\subseteq \Bc ) 
\big)
\notag\\&\leq 
\sum_{\Pc:\Pc\subseteq[N^{\prime}] , \Pc\not\subseteq \Bc}
H\big( V_{\Pc} \big)
\notag\\&\leq 
\sum_{\Pc:\Pc\subseteq[N^{\prime}] , \Pc\not\subseteq \Bc}
S_{\Pc}, \ \text{ where $S_{\Pc}=\log_{|\mathcal{X}|}(|\mathcal{V}_{\Pc}|)$}. 
\label{eq:composite ineqs}
\end{align}

In the following, we assume $|\Xc|$ is large enough such that $\{\log_{2}(|\mathcal{X}|) S_{\Pc} : \Pc\subseteq [N^{\prime}]\}$ are integers; this is so because the cardinality of the input alphabet does not affect the capacity region as argued in Remark~\ref{rem:|X| does not matter}. 
We choose the auxiliary random variables $(U_{i}: i\in [N^{\prime}])$ and $(V_{\Pc}: \Pc\subseteq[N^{\prime}])$ such that all the inequality leading to~\eqref{eq:composite ineqs} holds with equality for any $\Bc\subseteq[N^{\prime}]$, that is, we construct random variables $(V_{\Pc}: \Pc\subseteq[N^{\prime}])$ that are independent and uniformly distributed, where the alphabet of $V_{\Pc}$ has support of size $|\mathcal{V}_{\Pc}|=|\mathcal{X}|^{S_{\Pc}}=2^{\log_{2}(|\mathcal{X}|) S_{\Pc}}$.
With this choice of auxiliary random variables we show that Theorem~\ref{thm: novel index coding inner bound} reduces to Theorem~\ref{thm: thm2 composite coding}.

More precisely, let $U_{i}$, for $i\in [N^{\prime}]$, be an independent and equally likely binary vector of length $L_{i}$. 
For all $\Pc\subseteq[N^{\prime}]$, let $V_{\Pc}$ be a binary vector of length $\log_{2}(|\mathcal{X}|)  S_{\Pc}$ obtained as a linear code for the collection of bits in $(U_{i} : i\in\Pc)$.
We let $L_{i}=\sum_{\Pc\subseteq[N^{\prime}]:i\in\Pc} \log_{2}(|\mathcal{X}|) S_{\Pc}$ for all $i\in [N^{\prime}]$, and  divide the $L_{i}$ bits in $U_{i}$ into $2^{N^{\prime}-1}$ non-overlapping parts, where $U_{i}=(U_{i,\Pc}:\Pc\subseteq[N^{\prime}]:i\in\Pc)$ and $|U_{i,\Pc}|=\log_{2}(|\mathcal{X}|) S_{\Pc}$.
For each $\Pc\subseteq[N^{\prime}]$ where $|\Pc|>0$,  we let 
\begin{align}
V_{\Pc}=\underset{i\in\Pc}{\oplus}U_{i,\Pc}.\label{eq:Xp}
\end{align}
Now let us focus on a set $\Bc\subseteq[N^{\prime}]$. We have
\begin{align}
&
H\big((V_{\Pc}:\Pc\subseteq[N^{\prime}] ) 
\big|
(U_{i} : i\in \Bc) \big)\notag\\
&=  H\big( (V_{\Pc}:\Pc\subseteq[N^{\prime}], \Pc\not\subseteq \Bc\big|(U_{i} : i\in \Bc) \big) \notag\\
&=\negmedspace\negmedspace\negmedspace\negmedspace \sum_{j\in \left[2^{N^{\prime}}-2^{|\Bc|} \right]} \negmedspace\negmedspace\negmedspace\negmedspace\negmedspace\negmedspace H\big( V_{\Pc(\Bc,j)} \big|(U_{i} : i\in \Bc),V_{\Pc(\Bc,1)},\dots, V_{\Pc(\Bc,j-1)} \big), \label{eq:intermidiate}
\end{align}
where we randomly order the sets $\Pc\subseteq[N^{\prime}]$ where $\Pc\not\subseteq \Bc$, and we denote them by $\Pc(\Bc,1), \Pc(\Bc,2), \dots, \Pc(\Bc, 2^{N^{\prime}}-2^{|\Bc|})$.  We focus on one $j\in \left[2^{N^{\prime}}-2^{|\Bc|} \right]$,
\begin{subequations}
\begin{align}
& H\big( V_{\Pc(\Bc,j)} \big|(U_{i} : i\in \Bc),V_{\Pc(\Bc,1)},\dots, V_{\Pc(\Bc,j-1)} \big) \notag\\
&\geq  H\big( V_{\Pc(\Bc,j)} \big|(U_{i} : i\in \Bc), (U_{i,\Pc(\Bc,1)}:i\in  \Pc(\Bc,1)) ,\dots, \notag\\ &  (U_{i,\Pc(\Bc,j-1)}:i\in  \Pc(\Bc,j-1)) \big) \notag\\
&= H\big( V_{\Pc(\Bc,j)}  |(U_{i} : i\in \Bc) \big)\label{eq:independent 1}\\
&=H\big( V_{\Pc(\Bc,j)}   \big)\label{eq:independent 2}\\
&= \log_{|\mathcal{X}|}\big(2^{\log_{2}(|\mathcal{X}|) S_{\Pc(\Bc,j)}}\big)\label{eq:independent 3}\\
&= S_{\Pc(\Bc,j)},
\end{align}
\end{subequations}
where~\eqref{eq:independent 1} follows from the fact that $V_{\Pc(\Bc,j)} $ is independent of $ (U_{i,\Pc(\Bc,1)}:i\in  \Pc(\Bc,1)) ,\dots, (U_{i,\Pc(\Bc,j-1)}:i\in  \Pc(\Bc,j-1)) \big)$, ~\eqref{eq:independent 2} from that 
$\Pc(\Bc,j) \not\subseteq \Bc$,~\eqref{eq:independent 3} from that the bits in $ V_{\Pc(\Bc,j)}$ are i.i.d.. From~\eqref{eq:independent 2},~\eqref{eq:intermidiate} and~\eqref{eq:composite ineqs}, 
by our construction we can achieve
\begin{align}
H\big((V_{\Pc}:\Pc\subseteq[N^{\prime}] ) 
\big|
(U_{i} : i\in \Bc) 
\big)=\sum_{\Pc:\Pc\subseteq[N^{\prime}] , \Pc\not\subseteq \Bc}
S_{\Pc}. \label{eq:composite eqs}
\end{align} 


As a result, 
we have that
the bound in~\eqref{eq:H(X)<c} reduces to the one in~\eqref{eq:composite 4 decompression} 
by using~\eqref{eq:composite eqs} with $\Bc=\Ac_{j}$, and that
the bound in~\eqref{eq:wJ} reduces to the one in~\eqref{eq:composite 3 vJ}
by using~\eqref{eq:composite eqs} twice, once with $\Bc=\Ac \cup \Kc\setminus\Jc$ and once with $\Bc=\Ac \cup \Kc$, which is so because
\begin{align}
\kappa_{\Jc}
&=\sum_{\Pc:\Pc\subseteq[N^{\prime}] : \Pc\not\subseteq (\Ac \cup \Kc\setminus\Jc)} \!\!\!\!S_{\Pc}
 -\sum_{\Pc:\Pc\subseteq[N^{\prime}] : \Pc\not\subseteq (\Ac \cup \Kc)} \!\!\!\!S_{\Pc}
\\
&=\sum_{\Pc:\Pc\subseteq \Ac \cup \Kc : \Pc\cap\Jc\not=\emptyset} \!\!\!\!S_{\Pc}.
\end{align}
This concludes the proof. 

\bibliographystyle{IEEEtran}
\bibliography{IEEEexample}

\begin{thebibliography}{10}
\providecommand{\url}[1]{#1}
\csname url@samestyle\endcsname
\providecommand{\newblock}{\relax}
\providecommand{\bibinfo}[2]{#2}
\providecommand{\BIBentrySTDinterwordspacing}{\spaceskip=0pt\relax}
\providecommand{\BIBentryALTinterwordstretchfactor}{4}
\providecommand{\BIBentryALTinterwordspacing}{\spaceskip=\fontdimen2\font plus
\BIBentryALTinterwordstretchfactor\fontdimen3\font minus
  \fontdimen4\font\relax}
\providecommand{\BIBforeignlanguage}[2]{{%
\expandafter\ifx\csname l@#1\endcsname\relax
\typeout{** WARNING: IEEEtran.bst: No hyphenation pattern has been}%
\typeout{** loaded for the language `#1'. Using the pattern for}%
\typeout{** the default language instead.}%
\else
\language=\csname l@#1\endcsname
\fi
#2}}
\providecommand{\BIBdecl}{\relax}
\BIBdecl

\bibitem{dvbt2fundamental}
M.~A. Maddah-Ali and U.~Niesen, ``Fundamental limits of caching,'' \emph{IEEE
  Trans. Infor. Theory}, vol.~60, no.~5, pp. 2856--2867, May 2014.

\bibitem{decentralizedcoded}
------, ``Decentralized coded caching attains order-optimal memory-rate
  tradeoff,'' \emph{IEEE/ACM Trans. Networking}, vol.~23, no.~4, pp.
  1029--1040, Aug. 2015.

\bibitem{orderrandomJi2017}
M.~Ji, A.~M. Tulino, J.~Llorca, and G.~Caire, ``Order-optimal rate of caching
  and coded multicasting with random demands,'' \emph{IEEE Trans. Infor.
  Theory}, vol.~63, pp. 3923--3949, Jun. 2017.

\bibitem{ontheoptimality}
K.~Wan, D.~Tuninetti, and P.~Piantanida, ``On the optimality of uncoded cache
  placement,'' \emph{in 2016 IEEE Infor. Theory Workshop (first publicly
  available in Nov. 2015, at arXiv:1511.02256)}, 2016.

\bibitem{ourisitinnerbound}
------, ``On caching with more users than files,'' \emph{in 2016 IEEE Int.
  Symp. Inf. Theory (first publicly available in Jan. 2016, at
  arXiv:1601.06383)}, 2016.

\bibitem{ourd2dpaper}
C.~Yapar, K.~Wan, R.~F. Schaefer, and G.~Caire, ``On the optimality of d2d
  coded caching with uncoded cache placement and one-shot delivery,''
  \emph{available at arXiv:1901.05921}, Jan. 2019.

\bibitem{yener2018D2Dhetero}
A.~M. Ibrahim, A.~A. Zewail, and A.~Yener, ``Device-to-device coded caching
  with heterogeneous cache sizes,'' \emph{in IEEE Int. Conf. Communications
  (ICC)}, May 2018.

\bibitem{wan2017combinationouter}
K.~Wan, M.~Ji, P.~Piantanida, and D.~Tuninetti, ``Novel outer bounds for
  combination networks with end-user-caches,'' \emph{in IEEE Inf. Theory
  Workshop}, Nov. 2017.

\bibitem{parrinello2018sharedcache}
E.~Parrinello, A.~Unsal, and P.~Elia, ``Coded caching with shared caches:
  Fundamental limits with uncoded prefetching,'' \emph{arXiv:1809.09422}, Sep.
  2018.

\bibitem{approximatelyDS2018Attia}
M.~A. Attia and R.~Tandon, ``Approximately optimal distributed data
  shuffling,'' \emph{in IEEE Int. Symp. Inf. Theory}, Jul. 2018.

\bibitem{wang2016anewconverse}
C.-Y. Wang, S.~H. Lim, and M.~Gastpar, ``A new converse bound for coded
  caching,'' \emph{in Inf. Theory and Applications Workshop (ITA)}, Jan. 2016.

\bibitem{cachingwithlargeumberusers}
M.~M. Amiri, Q.~Yang, and D.~Gunduz, ``Coded caching for a large number of
  users,'' \emph{in IEEE Inf. Theory Workshop}, Sep. 2016.

\bibitem{exactrateuncoded}
Q.~Yu, M.~A. Maddah-Ali, and A.~Salman, ``The exact rate-memory tradeoff for
  caching with uncoded prefetching,'' \emph{in 2017 IEEE Int. Symp. Inf. Theory
  (first publicly available in Sep. 2016, at arXiv:1609.07817)}, 2017.

\bibitem{codedcachingviaintef}
C.~Tian and J.~Chen, ``Caching and delivery via interference elimination,''
  \emph{in IEEE Int. Symp. Inf. Theory}, pp. 830--834, Jul. 2016.

\bibitem{smallbufferusers}
Z.~Chen, ``Fundamental limits of caching: Improved bounds for small buffer
  users,'' \emph{IET Communications}, vol.~10, no.~17, Nov. 2016.

\bibitem{kuserstwofiles}
S.~Sahraei and M.~Gastpar, ``K users caching two files: An improved achievable
  rate,'' \emph{Annual Conference on Information Science and Systems (CISS)},
  Mar. 2016.

\bibitem{improveddelivery}
M.~M. Amiri and D.~Gunduz, ``Fundamental limits of caching: Improved delivery
  rate-cache capacity trade-off,'' \emph{IEEE Trans. Communications}, vol.~65,
  no.~2, pp. 806--815, Feb. 2017.

\bibitem{jesus2017fundamental}
J.~Gomez-Vilardebo, ``Fundamental limits of caching: Improved bounds with coded
  prefetching,'' \emph{available at arXiv:1612.09071}, Jan 2017.

\bibitem{improvedlower}
H.~Ghasemi and A.~Ramamoorthy, ``Further results on lower bounds for coded
  caching,'' \emph{in IEEE Int. Symp. Inf. Theory}, pp. 2319 -- 2323, Jul.
  2016.

\bibitem{criticaldatabase}
N.~Ajaykrishnan, N.~S. Prem, V.~M. Prabhakaran, and R.~Vaze, ``Critical
  database size for effective caching,'' \emph{National Conference on
  Communications (NCC)}, Feb. 2015.

\bibitem{ISIT2015outerbound}
A.~Sengupta, R.~Tandon, and T.~C. Clancy, ``Improved approximation of
  storage-rate tradeoff for caching via new outer bounds,'' \emph{in IEEE Int.
  Symp. Inf. Theory}, June. 2015.

\bibitem{demandtype}
C.~Tian, ``Symmetry, demand types and outer bounds in caching systems,''
  \emph{in IEEE Int. Symp. Inf. Theory}, pp. 825--829, Jul. 2016.

\bibitem{improvedconverse2017Wang}
C.-Y. Wang, S.~S. Bidokhti, and M.~Wigger, ``Improved converses and gap-results
  for coded caching,'' \emph{IEEE Trans. Infor. Theory}, vol.~64, no.~11, pp.
  7051--7062, Jul. 2018.

\bibitem{CoverThomas2ndEdition}
T.~M. Cover and J.~A. Thomas, ``Elements of information theory,'' \emph{2nd ed.
  New York, NY, USA: Wiley-interscience}, 2006.

\bibitem{benefitofcache2017Shirin}
S.~S. Bidokhti, M.~Wigger, and A.~Yener, ``Benefits of cache assignment on
  degraded broadcast channels,'' \emph{in IEEE Int. Symp. Inf. Theory}, Jun.
  2017.

\bibitem{yu2017characterizing}
Q.~Yu, M.~A. Maddah-Ali, and A.~Salman, ``Characterizing the rate-memory
  tradeoff in cache networks within a factor of 2,'' \emph{in IEEE Int. Symp.
  Inf. Theory}, Jun. 2017.

\bibitem{novelouterwan2017}
K.~Wan, M.~Ji, P.~Piantanida, and D.~Tuninetti, ``Novel outer bounds with
  uncoded cache placement for combination networks with end-user-caches,''
  \emph{in IEEE Inf. Theory Workshop (ITW)}, Nov. 2017.

\bibitem{onthecapacityindex}
F.~Arbabjolfaei, B.~Bandemer, Y.-H. Kim, E.~Sasoglu, and L.~Wang, ``On the
  capacity region for index coding,'' \emph{in IEEE Int. Symp. Inf. Theory},
  Jul. 2013.

\bibitem{hanspaper}
T.~S. Han, ``The capacity region of general multiple-access channel with
  certain correlated sources,'' \emph{Inf. Contr.}, vol.~40, no.~1, pp. 37--60,
  1979.

\bibitem{slepianwolf}
D.~Slepian and J.~K. Wolf, ``Noiseless coding of correlated information
  sources,'' \emph{IEEE Trans. Infor. Theory}, vol.~19, no.~4, pp. 471--480,
  Jul. 1973.

\bibitem{elgamalkim}
A.~E. Gamal and Y.-H. Kim, ``Network information theory,'' \emph{Cambridge
  University Press}, 2011.

\bibitem{kai-defense}
K.~Wan, ``Fundamental limits of cache-aided shared-link broadcast networks and
  combination networks,'' \emph{Ph.D. thesis}, Jun. 2018, available at
  https://tel.archives-ouvertes.fr/tel-01842269/.

\bibitem{birk1998informedsource}
Y.~Birk and T.~Kol, ``Informed source coding on demand (iscod) over broadcast
  channels,'' \emph{in Proc. IEEE Conf. Comput. Commun.}, p. 1257–1264, 1998.

\bibitem{fundamentalindexcoding}
F.~Arbabjolfaei and Y.-H. Kim, ``Fundamentals of index coding,''
  \emph{Foundations and Trends in Communications and Information Theory},
  vol.~14, no. 3-4, pp. 163--346, Oct. 2018.

\bibitem{Lexicographic}
A.~Blasiak, R.~Kleinberg, and E.~Lubetzky, ``Lexicographic products and the
  power of non-linear network coding,'' \emph{in IEEE 52nd Annu. Symp. Found.
  Comput. Sci.}, pp. 609--618, Oct. 2011.

\bibitem{liu2017ondistributed}
Y.~Liu, P.~Sadeghi, F.~Arbabjolfaei, and Y.-H. Kim, ``On the capacity for
  distributed index coding,'' \emph{in IEEE Int. Symp. Inf. Theory}, Jun. 2017.

\bibitem{nonunique}
B.~Bandemer, A.~E. Gamal, and Y.~H. Kim, ``Simultaneous nonunique decoding is
  rate-optimal,'' \emph{in Proc. 50th Allerton Conf.}, pp. 9--16, Oct. 2012.

\end{thebibliography}

	\begin{IEEEbiographynophoto}
						{Kai Wan} (S '15 -- M '18)
						received  the M.Sc. and Ph.D. degrees in Communications from Universit{\'e} Paris Sud-CentraleSup{\'e}lec, France, in 2014 and 2018.  He is currently a post-doctoral researcher with the Communications and Information Theory Chair (CommIT) at Technische Universit\"at Berlin, Berlin, Germany. His research interests include coded caching,  index coding, distributed storage, wireless communications, and distributed computing.
					\end{IEEEbiographynophoto}
					
										\begin{IEEEbiographynophoto}{Daniela Tuninetti}  (M '98 -- SM '13)
 is currently a Professor within the Department of Electrical
and Computer Engineering at the University of Illinois at Chicago (UIC),
which she joined in 2005. Dr. Tuninetti got her Ph.D. in Electrical Engineering
in 2002 from ENST/T{\'e}l{\'e}com ParisTech (Paris, France, with work done at the
Eurecom Institute in Sophia Antipolis, France), and she was a postdoctoral
research associate at the School of Communication and Computer Science
at the Swiss Federal Institute of Technology in Lausanne (EPFL, Lausanne,
Switzerland) from 2002 to 2004. Dr. Tuninetti is a recipient of a best paper
award at the European Wireless Conference in 2002, of an NSF CAREER
award in 2007, and named University of Illinois Scholar in 2015. Dr. Tuninetti
was the editor-in-chief of the IEEE Information Theory Society Newsletter
from 2006 to 2008, an editor for IEEE COMMUNICATION LETTERS from
2006 to 2009, for IEEE TRANSACTIONS ON WIRELESS COMMUNICATIONS
from 2011 to 2014; and for IEEE TRANSACTIONS ON INFORMATION
THEORY from 2014 to 2017. She is currently a distinguished lecturer for the
Information Theory society. Dr. Tuninetti's research interests are in the
ultimate performance limits of wireless interference networks (with special
emphasis on cognition and user cooperation), coexistence between radar and
communication systems, multi-relay networks, content-type coding, cache-aided
systems and distributed private coded computing.
					\end{IEEEbiographynophoto}

											\begin{IEEEbiographynophoto}{Pablo Piantanida}
(SM '16) received both B.Sc. in Electrical Engineering and the M.Sc (with honors) from the University of Buenos Aires (Argentina) in 2003, and the Ph.D. from Universit{\'e} Paris-Sud (Orsay, France) in 2007. Since October 2007 he has joined the Laboratoire des Signaux et Syst{\`e}mes (L2S), at CentraleSup{\'e}lec together with CNRS (UMR 8506) and Universit{\'e} Paris-Sud, as an Associate Professor of Network Information Theory. He is currently associated with Montreal Institute for Learning Algorithms (Mila) at Universit{\'e} de Montr{\'e}al, Quebec, Canada.  He is an IEEE Senior Member and serves as Associate Editor for IEEE Transactions on Information Forensics and Security. He served as General Co-Chair of the 2019 IEEE International Symposium on Information Theory (ISIT). His research interests lie broadly in information theory and its interactions with other fields, including multi-terminal information theory, Shannon theory, machine learning and representation learning, statistical inference, cooperative communications, communication mechanisms for security and privacy.
					\end{IEEEbiographynophoto}

\end{document}